\documentclass[pra,aps,final,twocolumn,nofootinbib,showpacs]{revtex4-1}
\usepackage{epsfig}
\usepackage{latexsym}
\usepackage{xspace}
\usepackage{hyperref}
\usepackage[latin2]{inputenc}
\usepackage{indentfirst}
\usepackage{enumerate}
\usepackage{color}
\usepackage{colordvi}

\usepackage{amsmath}
\usepackage{amssymb}
\usepackage[english]{babel}
\usepackage{url}

\def\be{\begin{equation}}
\def\ee{\end{equation}}
\def\bea{\begin{eqnarray}}
\def\eea{\end{eqnarray}}
\def\l{\label}

\def\d{\mbox{d}}

\def\siml{\;\hbox{\kern.1em \lower.7ex \hbox{$\sim$} \kern-1.12em
 \raise.5ex \hbox{$<$} \kern.1em}}
\def\simg{\;\hbox{\kern.1em \lower.7ex \hbox{$\sim$} \kern-1.12em
 \raise.5ex \hbox{$>$} \kern.1em}}

\def\siml{\;\hbox{\kern.1em \lower.7ex \hbox{$\sim$} \kern-1.12em
 \raise.5ex \hbox{$<$} \kern.1em}}
\def\simg{\;\hbox{\kern.1em \lower.7ex \hbox{$\sim$} \kern-1.12em
 \raise.5ex \hbox{$>$} \kern.1em}}

\begin{document}

\title{
  Particle-number fluctuations
  near the critical point of nuclear matter}

\author{A.G.\ Magner}
\affiliation{Institute for Nuclear Research NASU, 03028 Kyiv, Ukraine}
\affiliation{Cyclotron Institute, Texas A\&M University,
  College Station, Texas 77843, USA}
\author{S.N.\ Fedotkin}
\affiliation{Institute for Nuclear Research NASU, 03028 Kyiv, Ukraine}
\author{U.V.\ Grygoriev}
\affiliation{Institute for Nuclear Research NASU, 03028 Kyiv, Ukraine} 
\affiliation{University of Groningen, Van Swinderen Institute for
Particle Physics and Gravity, 9747 AG, Groningen, Netherlands}

  \begin{abstract}
The equation of state with 
quantum statistics corrections
is used for particle number fluctuations $\omega$ of
isotopically symmetric
nuclear matter with interparticle 
van der Waals and Skyrme
local density interactions.
The fluctuations, $\omega\propto 1/\mathcal{K}$,
are analytically derived through  
the isothermal incompressibility
   $\mathcal{K}$
at first order 
 over a small  quantum-statistics parameter.
 Our approximate analytical results appear to be in  
  good
  agreement with the results of
  accurate numerical calculations.
    These results are 
      also close 
  to those obtained by using  
  more accurate 
 Tolman and Rowlinson expansions of the incompressibility $\mathcal{K}$ near
 the critical point.  A more general formula
         for fluctuations $\omega$, improved  at the critical point,
     was obtained for a finite particle-number average $\langle N \rangle$ by 
     neglecting, for simplicity,
     small quantum statistics
     effects. It is shown that for a large dimensionless parameter,
     $\alpha \propto
     \mathcal{K}^2\langle N \rangle/\mathcal{K}^{\prime\prime} $,
     where $\mathcal{K}^{\prime\prime}$ is the second derivative of the
     incompressibility
     $\mathcal{K}$ as function of the average particle density $n$,
     far from the critical point
     ($\alpha \gg 1$),
     one finds the traditional asymptote, $\omega\propto 1/\mathcal{K}$,
     for the
     fluctuations $\omega$.
     For a small parameter,
     $\alpha \ll 1$, near the critical
     point, where $\mathcal{K}=0$ and $\alpha=0$,
     one obtains another asymptote of $\omega$. These fluctuations, having a
     maximum near the
     critical point  as function of the average density $n$,
     for finite values of $\langle N \rangle$
     are finite and relatively small,  
                           in contrast to the results of the
     traditional calculations.
  
\end{abstract}
%

\maketitle

\section{Introduction}
\l{sec-introd}

Many works have been devoted to studying the properties of 
    nuclear systems with
strongly interacting
 particles;
see, e.g.,
Refs.~\cite{bethe,migdal,MS69,BD72,RS80,BG85,BHR03,KS20}. 
Realistic versions of the nuclear matter equation of state 
    include both
attractive and repulsive forces
between
particles. 
Thermodynamical behavior of this matter
leads to the liquid-gas first-order phase transition
which ends at the critical point; see Refs.~\cite{LLv5,huang,AB00} and,
for special emphasis on finiteness of nuclear systems in 
multifragmetation reactions, Refs.~\cite{AB08,AB10}.
Experimentally, the presence of the liquid-gas phase transition
in nuclear matter was  reported and then analyzed
in numerous papers
(see, e.g.,
Refs.~\cite{ex-1,ex-2,ex-3,ex-4,ex-5,ex-5a,ex-6,AB08,AB10}).
Critical points in different systems of nuclear matter were studied in
many theoretical works; see,
e.g., recent
 Refs.~\cite{vova,satarov,roma1}.
 These works, which are mainly based on
 the proposed  van der Waals (vdW) and effective Skyrme
 local density (SLD) approaches  
for the equation of state
accounting for 
quantum statistics (QvdW and QSLD) 
\cite{marik,vova},
were used to describe the
properties of 
nuclear matter. Also,   
extensions for many-component systems, 
and applications
    to the
    fluctuation calculations (see, e.g.,
    Refs.~\cite{roma,satarov1,roma2,oleh20,St21-1,St21-2,Kuzn21})
 were suggested for different thermodynamical averages
\cite{TR38,RJ58,LLv5,TK66,KG67,IA71,BR75,AC90,An87,Sa99,ZR88,ZM02,KW04,AB10}.

 The role and size of the effects of 
quantum statistics was studied analytically for nuclear matter, and also
for pure neutron and pure $\alpha$-particle matter, in 
    Refs.~\cite{FMG19,FMG22}.
    An analytical expression for 
    dependence of
the critical point parameters on
the particle mass $m$,
degeneracy factor $g$, and the 
QvdW and 
QSLD 
 interaction constants $a$, $b$ (or their matrices)
 for the vdW and those including $\gamma$
for the 
 SLD 
 was derived in Refs.~ \cite{FMG19} and \cite{FMG22}, respectively.
In particular, the analytical approach  of Ref.~\cite{FMG19}
was extended \cite{FMG22}
to the effective simple SLD approach 
\cite{AV-15,satarov0,satarov}; see also review articles \cite{La81}.
 This approach
is related to the Skyrme forces through the potential part of the
local energy-density functional.
Our consideration 
was restricted to relatively small temperature, 
$T \siml 30$~MeV, and not too large
particle density. On the other hand, the temperature $T$
should be sufficiently large to satisfy the smallness of the
quantum statistics parameter \cite{LLv5}. Within these restrictions, the
number of nucleons can be determined 
by
a conservation rule, 
and the chemical potential of 
such systems is regulated by the particle number density of
 nuclear matter.
An extension of the formulation to  
fully relativistic hadron resonances
    in a gas system of baryons and antibarions with vdW two-body 
interactions 
was considered in 
Ref.~\cite{VGS-17}.
Applications of this extended model to the
net baryon number fluctuations
in relativistic nucleus-nucleus
collisions was developed in Refs.~
\cite{roma,VJGS-18,satarov1,roma2,oleh20,St21-1,St21-2,Kuzn21}.
We do not include the Coulomb forces
and make no distinction between protons and neutrons
(both these particles are referred to as nucleons).
In addition, under these restrictions the
 nonrelativistic treatment becomes very accurate
and is adopted in our studies.

In the present work we are going to apply
the same analytical method as for derivations of the equation of state,
    taking
    into account the quantum statistics effects in terms of a few
    first corrections
within the QvdW and 
 QSLD models
  (see  Refs.~\cite{FMG19,FMG22}),
to analyze the
particle number fluctuations near the critical point of
 nuclear matter; see also Ref.~\cite{FMG20}.  
    Different analytical and numerical approximations to these fluctuations
    might be helpful to determine ranges of their applicability.
  As shown, e.g., in
  Ref.~\cite{ZM02},
  the expression of the particle
   number fluctuation, $\omega$,
    in terms
    of the susceptibility, and then, through the
    incompressibility, $\omega \propto 1/\mathcal{K}$, was derived
    from the 
    original definition for $\omega$ in terms of
    the moments of the Gibbs distribution, 
        averaged over phase space
    variables, by assuming the smallness of the
    fluctuations. However, the limit of this traditional
    expression to the critical point, where
    $\mathcal{K}=0$, is obviously divergent. Divergences
    of fluctuations $\omega$ near the critical point are incompatible
    with a more accurate
    equation of state, accounting for interparticle interaction, as a
    relation between statistically averaged characteristics of the
     desired system, which are determined
    up to these fluctuations \cite{RJ58,TK66,KG67,ZM02}. The main critique
     of these divergences is that they are due to 
    highly
    idealized assumptions
     (e.g., a mean-field approach up to
        particle correlations in the infinite system \cite{huang,AB00}).
    These assumptions were used in 
     the derivations of fluctuations $\omega$ from the moments of the
     Gibbs distribution
    over particle number $N$ in the grand canonical ensemble. 
    The traditional
    expression for 
    the fluctuation ($\omega\propto 1/\mathcal{K}$)
    in terms of the
    incompressibility
    $\mathcal{K}$
    fails near 
    the critical point. Some suggestions to overcome the divergence problem
    for the fluctuations, taking into account the particle correlations,
    can be found in Ref.~\cite{KG67}. 
    For the vdW problem, one can find other specific semianalytical
    suggestions
    in Ref.~\cite{KW04}.
    We will apply 
    another statistical
    approach \cite{Sm08,Ei10} based on 
     expansion of the free energy $F$ over powers of
        a small difference of the
     particle number density $\rho$ and its average $n$ for
     a given temperature $T$. We are going to use
     explicitly the
     statistical 
     Gibbs distribution  averaged, however, in
       phase space for calculations of the particle number density
    dispersion and the corresponding
    fluctuation $\omega$;  see also
    Refs.~\cite{TR38,RJ58,TK66}. As is well known (see, e.g.,
    \cite{TR38,TK66}), the expansion of $F$ at second order
    leads to the traditional expression for the particle number fluctuations,
    $\omega \propto 1/\mathcal{K}$. Taking into account only
    fourth-order terms,  and
    neglecting the second-order ones in 
        a very close vicinity to the critical point, one has
     several improved
    results derived
    in Refs.~\cite{RJ58,TK66}. We will take into account 
    both fourth- and second-order terms and obtain
    a more general result for these
    fluctuations, accurately determining the two
    asymptotes far from and close to
    the critical point
        for a finite average particle number. 
   In order to compare in detail
   a more general asymptote and the two above-mentioned asymptotes
       near the critical point,
    we take
     the phenomenological vdW and SLD density-dependent
    interactions as well-known examples. In this way, we will neglect
  the 
  quantum statistics effects which are not very important
  for fluctuations, in contrast to the critical point calculations.
   Notice that the order parameter $\rho - n$ in our approach
      is essentially
      different from $\rho-n_c$, where $n_c$ is the critical value of the
      average particle number density $n$, in the
       Landau    local fluctuation
     theory.
       Therefore, in our mean-field approach,
       up to statistical correlations,
       it is possible to cross the critical point with
         finite fluctuations
      by changing a dimensionless parameter,
      $\alpha \propto K^2 \langle N \rangle/ n^2T K^{\prime\prime}$
      to zero. In this approach,
  the second derivative $K^{\prime\prime}$ of the isothermal incompessibility
  $K$ is assumed to be not zero at the critical point ($K=0$) for a finite
  average of the particle numbers $N$, $\langle N \rangle$. 
    Otherwise, we will
    need to use an expansion of the free energy $F$ up to high order terms.
  The parameter
  $\alpha$ is a measure of the effective distance from a critical point, which
  depends on the interparticle interaction and average particle number
  $\langle N \rangle$.
  Thus, we will have
  a transition from large effective parameter $\alpha$ of the
  traditional formula for the fluctuations $\omega$ to the 
   Rowlinson formula 
      \cite{RJ58} 
  at small $\alpha$, which is local near the
  critical point, within the
  Smoluchowski and Einstein fluctuation theory
  \cite{Sm08,Ei10}.  To some extent, that is a more general approach than
  the classical Landau fluctuation
  theory. 
      The fluctuation calculations based on the
      statistically averaged level density 
       obtained analytically in 
          Refs.~\cite{MS21npa,MS21prc,MS21ijmpe,MS22FLT} will be 
     adopted to determine the particle
     number fluctuations for finite nuclear systems in  a 
     forthcoming work.

     The paper is organized as follows.
         In Sec. 
         ~\ref{sec-2} we review some general relationships of the
         statistical physics used in our derivations.
         In Sec.~\ref{sec-3}, the known
           analytical derivations and results
     for the classical
     fluctuations are presented following Refs.~\cite{LLv5,TR38,RJ58,TK66}.
     We apply them for the
traditional 
 analytical method of
 fluctuation calculations in terms of the incompressibility
  for the vdW and SLD
 interparticle interactions in Sec.~\ref{sec-4}.
Then, in Sec.~\ref{sec-5}, we present the improved derivations 
for the fluctuation calculations 
based on the statistically averaged 
Gibbs distribution. The same
vdW and SLD interaction models are taken,
    as simple exemplary cases.
     All obtained results are discussed in Sec.~\ref{sec-6}.
    We
    compare our analytical traditional
    calculations with those
 modified 
by Tolman \cite{TR38} and Rowlinson
\cite{RJ58}, and with numerical calculations 
    carried out by Gorenstein and his
    collaborators in Refs.~\cite{marik,vova,roma}.
 The same parameters of simple
    phenomenological interactions,
    vdW and SLD, are used for the comparison 
        between 
    the results of
    Secs.~\ref{sec-4} and \ref{sec-5}.
  These results are summarized in
 Sec. \ref{sec-sum}. Some details of our derivations are presented
 in Appendices \ref{appA}-\ref{appF}.
 Pecularities of the classical fluctuations
 in terms of the first- and high-order susceptibilities and
 incompressibilities of nuclear matter are discussed in
 Appendices \ref{appA} and \ref{appB}, respectively. 
In Appendix \ref{appC}, the analytical results for the critical point are
reviewed
for  the case of the QvdW and  QSLD
approaches taking into account the
    quantum statistics corrections 
    to the vdW (Ref.~\cite{FMG19,FMG22}) and
    SLD \cite{FMG22} models. 
        In Appendices~\ref{appD} and \ref{appE}, following
        Ref.~\cite{TR38,RJ58} we present
        some details of the derivations of the general improved
        fluctuation formula and its asymptote near the critical point,
        respectively.  Appendix~\ref{appF} is devoted
            to our approach following the
    Tolpygo classical fluctuation theory \cite{TK66}.

         \section{General points}
         \l{sec-2}

For calculations of classical fluctuations of the particle
numbers, $\omega$, within the grand canonical ensemble, 
one can start with the particle number average \cite{ZM02,LLv5}

\be\l{avpartnumb}
\langle N \rangle=\sum^{}_{N} N \int W_{\rm eq}^{(N)}({\bf q},{\bf p}; T,\mu,V)
\d \Gamma^{}_N~.
\ee
Here, $W_{\rm eq}^{(N)}({\bf q},{\bf p}; T, \mu, V)$
is the Gibbs
distribution function
 of 
    phase space variables ${\bf q},{\bf p}$;  
     $\d \Gamma^{}_N=
\d {\bf q}\d {\bf p}$ for a given particle number $N$
(normalized as usually for a classical system).
   Other variables,
  $T$, $\mu$, and $V$, are the temperature, 
     chemical potential, and system volume in the grand canonical ensemble,
     respectively.
The
Gibbs probability distribution $W^{(N)}_{\rm eq}$ can be written in terms
of the classical Hamiltonian
$H_N({\bf q},{\bf p})$ as  
\bea\l{distfun}
&W_{\rm eq}^{(N)}({\bf q},{\bf p}; T, \mu,V)=\frac{1}{\mathcal{Z}(T,\mu,V)}
\nonumber\\
&\times \exp\left\{-\left[H_N({\bf q},{\bf p}) -\mu N\right]/T\right\}~.
\eea
The Hamiltonian
$H_N({\bf q},{\bf p})$ is the basic part 
also for the
normalization  factor,   
\be\l{partfun}
\hspace{-0.3cm}\mathcal{Z}(T,\mu,V)=
\!\sum^{}_N 
\!\int\!
\d\Gamma^{}_{N}\exp\left\{-\left[H_N({\bf q},{\bf p})
  \!-\!\mu N\right]/T\right\}.
\ee
Thus, the partition function $\mathcal{Z}(T,\mu,V)$ 
     obeys the
normalization condition
for the distribution
$ W^{(N)}_{\rm eq}$. 
This distribution,
averaged below over the
phase space variables ${\bf q}$ and ${\bf p}$ for a given particle
number $N$, will be denoted
as $\overline{W}^{(N)}_{\rm eq}$
  (the line above the quantity means
averaging only over the phase space variables). In addition,
  the averaging
over 
particle numbers $N$, as in Eq.~(\ref{avpartnumb}), along with averaging over
the phase space variables,
will be shown by angle brackets,
$\langle W^{(N)}_{\rm eq}\rangle$;
see Eq.~(\ref{distfun}) and Ref.~\cite{ZM02}.
Using these notations, for the 
    classical entropy $S(T,\mu,V)$ one has
\bea\l{entropy}
&S(T,\mu,V)= -\langle \ln W^{(N)}_{\rm eq}\rangle =
\frac{1}{T}\left[\langle H \rangle\right.\nonumber\\
 & -\left.\mu \langle N\rangle +
T \ln \mathcal{Z}(T,\mu,V)\right]~.
\eea
We may now introduce 
the equilibrium thermodynamical potential,
$\Omega(T,\mu,V)$, for the grand canonical ensemble with
the help of the relationship
\be\l{Omega}
\Omega=F-\mu \langle H \rangle=U-TS-\mu \langle H \rangle,
\quad U=\langle H\rangle~.
\ee
The free energy $F(T,N,V)$ of the canonical ensemble
is considered as function of
the temperature $T$,
 particle number 
$N$, and system volume $V$,  
\be\l{F}
F=U-TS~. 
\ee
Then, the 
chemical potential $\mu$ can be defined in terms of the free energy $F$,
$\mu=(\partial F/\partial N)_T$ in the canonical ensemble.
From Eq.~(\ref{Omega})
one finds the standard expression \cite{LLv5} for a thermodynamical
potential $\Omega$
of the grand canonical ensemble in terms of the partition function
$\mathcal{Z}$
[Eq.~(\ref{partfun})],
\be\l{OmZ}
\Omega(T,\mu,V)=-T \ln \mathcal{Z}(T,\mu,V)~.
\ee
 For intensively large 
    systems,
    one can consider a local density
    of the thermodynamic potential $\Omega$ per unit of 
    volume $V$, i.e., the pressure $P(T,n)$
    \cite{LLv5}. 
    For such intensive systems in the grand canonical ensemble, one has
    $\Omega=-V P(T,n)$, where 
   we can neglect the explicit
    volume dependence
    of the pressure, $P(T,n)$. This dependence is realized
    only through the averaged
    particle number density $n=N/V$.
    The equation of state, $P=P(T,n)$, can be found
    through the explicit expression for $P(T,n)$ as a function of temperature
    $T$ and particle number density $n$. This takes place, e.g., if we
    can neglect the surface part of the system
    pressure (the capillary pressure
    due to the surface tension) with respect to its volume part.

\section{Classical fluctuations}
\l{sec-3}

As mentioned in the Introduction, the derivation of 
   the expression 
for the particle number fluctuations $\omega$ in terms of the
incompressibility $\mathcal{K}$ from the moments of 
a mean
Gibbs distribution $\overline{W}^{(N)}_{\rm eq}$ is questionable near the
critical point (CP), where $\mathcal{K}=0$. Indeed, the
assumption of
smallness of $\omega$ in 
this derivation  
fails near the CP (see, e.g., Refs.~\cite{TR38,RJ58,TK66}).
Therefore, to clarify the 
behavior of fluctuations near the CP, one has to
    consider more accurately
the derivations within the classical fluctuation theory \cite{Sm08,Ei10}
beginning from the 
dispersion (squared) of the particle number distribution:
\be
\mbox{D}_N=
\langle \left (\Delta N \right)^2\rangle=
\langle N^2 \rangle -
                \langle N \rangle^2,
                  \l{FL-gendef}
\ee
where $\Delta N=N-\langle N \rangle$ is 
   the deflection of the
    particle number
$N$ from 
its average
$\langle N \rangle$. Averages $\langle N^\kappa \rangle$ are the
mean $\kappa$ moments of the distribution function $\overline{W}^{(N)}_{\rm eq}$
 (see the previous section), 
    \bea\l{momGN}
    & \langle N^\kappa \rangle =
    \int N^\kappa \overline{W}^{(N)}_{{\rm eq}}~ \d N,\nonumber\\
  &  \int \overline{W}^{(N)}_{{\rm eq}}~ \d N=1,~~~\kappa=1,2~.
    \eea
   The particle number dispersion 
$\mbox{D}_N$ [Eq.~(\ref{FL-gendef})] 
    can be expressed in terms of
    these two moments, $\langle N \rangle $ and $\langle N^{2} \rangle$,
    of the Gibbs distribution $\overline{W}^{(N)}_{\rm eq}$, averaged
    over the phase space variables;
    see above.
    For intensive systems,  it is convenient to calculate first
    the dispersion $\mathcal{D}_\rho$ 
    of the particle number density fluctuations, e.g.,
    in units of $n^2$ for the dimensionless reason,
\be\l{FL-MM}
\frac{\mathcal{D}_\rho}{n^2} =  
\frac{\langle \left (\Delta \rho \right)^2\rangle}{n^2}=
  \frac{\langle \rho^2 \rangle -
  \langle \rho \rangle^2}{n^2},
\ee
where  $\Delta \rho=\rho-\langle \rho \rangle$ is 
   the deflection of the 
     particle number density 
$\rho$ from 
its average
$\langle \rho \rangle=n$    
      (see  
    Ref.~\cite{TR38}).

    We will discuss
    the normalization of the
    dispersion $\mbox{D}_N$ in terms of the particle number
    fluctuations
    later. First, we will show
    that the dispersions (\ref{FL-gendef}) and (\ref{FL-MM}) are
    significantly different
    by order of the power of $<N>$, linear and quadratic, far and near
    the critical point, respectively. 
    
 The angle brackets in Eq.~(\ref{FL-MM}) are defined 
by
\be
\langle \rho^\kappa \rangle =
\int_{0}^{\rho_{\rm up}}  \mbox{d} \rho~\rho^\kappa~W(\rho),\quad
  \int_{0}^{\rho_{\rm up}}
  \mbox{d} \rho~
  W(\rho)=1~, 
    \l{avN}
\ee
 where again
 $\kappa=1,2$ and $\rho_{\rm up}$ is the upper limit of the integration over
 particle number density $\rho$. Notice that for the vdW
 interparticle interaction
 one has a restriction to the upper limit $\rho_{\rm up}$ of this
 integration by
   its value $1/(3b)$, where $b$ is the volume exclusion parameter
   \cite{LLv5,FMG19,FMG22}, while $\rho_{\rm up}=\infty$ for the SLD
   interaction
   case.  We will expand approximately
 $\rho_{up}$ to the infinity in all following normalization integrals.
 The
    probability distribution function $W(\rho)$ in Eq.~(\ref{avN})
    can be approximated 
    by \cite{TR38}
\be\l{Wrho}
W(\rho)=W^{(0)}\exp\left[-\frac{F(\rho)-F(n)}{T}\right],
\ee
where $W^{(0)}$ is the normalization constant,
\be\l{W0}
W^{(0)}=\left\{\int_{0}^\infty \mbox{d} \rho~
\exp\left[-\frac{F(\rho)-F(n)}{T}\right]
\right\}^{-1}.
\ee
We omit the temperature variable $T$ in the free energy $F$
because it is a constant
in all our following derivations.

  Following the ideas of Smoluchowski and Einstein
  (see Refs.~\cite{Sm08,Ei10,TR38}),
for small fluctuations,
        the free energy 
        $F(\rho)$ 
            for an intensive system
        can be expanded in powers of a
        difference between the particle number density, $\rho$, and
        its statistical average, $\langle \rho \rangle=n$.
     Up to fourth-order terms for the fixed
    temperature $T$, one writes
    \bea\l{Fexp}
    &F(\rho)=F(n) +
    \frac12 \left(\frac{\partial^2 F}{\partial \rho^2}\right)^{}_{\rho=n}
    \left(\rho-n\right)^2 \nonumber\\
    & +\frac{1}{24}
    \left(\frac{\partial^4 F}{\partial \rho^4}\right)^{}_{\rho=n}
    \left(\rho-n\right)^4 + \ldots~.
    \eea
    At fourth order, as will be used below, one writes   
    \bea\l{DF}
    &\Delta\{\rho\}\equiv F(\rho)-F(n) =
    \frac12 \left(\frac{\partial^2 F}{\partial \rho^2}\right)^{}_{\rho=n}
    \left(\rho-n\right)^2 \nonumber\\
    & +\frac{1}{24}
    \left(\frac{\partial^4 F}{\partial \rho^4}\right)^{}_{\rho=n}
    \left(\rho-n\right)^4. 
    \eea
    We introduced here the functional $\Delta\{\rho\}$ of the particle
    number density $\rho$.
    The first- and third-order terms in these expansions
    can be put to zero because
    our system is considered to be at  
       statistical equilibrium  with a minimum of
           the free energy $F(\rho)$ 
             and there are no external fields under consideration. We will
           assume also that
           the fourth-order terms dominate over high order terms.
           As shown in Refs.~\cite{huang,AB00,AB10}, a little more
           complexly
           but still analytically, six-order terms
           can be also taken into account for study of the
           tricritical point.
       Some appearing constants could be included
    in the normalization factor $W^{(0)}$;
    see Eqs.~(\ref{Wrho}) and (\ref{W0}). Using the 
    expansion of the
    free energy $F(\rho)$, Eq.~(\ref{Fexp}) at fourth order,
    one can immediately rewrite the probability distribution
    (\ref{Wrho}) as
    \bea\l{W4}
    &  W_4(\rho) =W^{(0)}_4
   \exp\left[-\frac{F_2}{2T}\left(
     (\rho-n)^2\right.\right.\nonumber\\
   &  +\left.\left.
      \frac{F_4}{12}
      (\rho-n)^4\right)
       \right]~,
    \eea
    where
    \bea\l{norm4}
  &  W^{(0)}_4=\left\{\int_0^\infty \mbox{d} \rho~
    \exp\left[-\frac{F_2}{2T}\left(
      (\rho-n)^2\right.\right.\right.
      \nonumber\\
      &+\left.\left.\left.
      \frac{F_4}{12 T}
      (\rho-n)^4\right)
      \right]\right\}^{-1},
    \eea
    and $F_2$ and $F_4$ are derivatives of the free energy $F$ given by
    Eq.~(\ref{Fm}).
    Indeed, according to
        Eqs.~(\ref{W4}) and (\ref{norm4}), one has the normalization
        condition,  
    \be\l{normcond}
\int_0^\infty W_m(\rho)\mbox{d}\rho=1, ~~~~m=2,4~. 
\ee
     Therefore, 
      assuming that
          the second-order correction in the  
         expansion (\ref{Fexp}) for the free energy $F$
         is relatively large with respect to high order terms,
            one can neglect all other fourth- and high-order terms in
             Eqs.~(\ref{W4}) and (\ref{norm4}),
    \be\l{W2}
    W_2(\rho) =
    W^{(0)}_2\exp\left[
      -\frac{F_2}{2T}
      (\rho-n)^2
      \right],
    \ee
    where
    \be\l{norm2}
    \hspace{-0.2cm}W^{(0)}_2=\left\{\int_0^\infty \mbox{d} \rho~\exp\left[
      -\frac{F_2}{2T}
      (\rho-n)^2
        \right]\right\}^{-1}.
    \ee
             Thus, from Eq.~(\ref{DF}) one finds
             \footnote{Notice that the distribution $W_2$,
             Eq.~(\ref{W2}),
             can be obtained also starting from the
             Gibbs expression, $\propto \exp(S)$, where $S$ is the entropy (see
             Ref.~\cite{LLv5}).  Expanding the entropy near the
                 statistical equilibrium,
                 $(\partial S/\partial \rho)_{\rho=n}=0$,
                 one has
             a probability distribution 
             of the same Gaussian form.
            We use here the  definitions of the entropy $S$
              and free energy $F$, 
              through the partition function $\mathcal{Z}$
               [see Eq.~(\ref{F})].
                  In addition, one can take into account that the high-order
              (second and higher)
             derivatives 
             of the entropy $S$
             at the equilibrium are identical to those of the free
             energy $F$ over the particle number density $\rho $
             for a constant temperature $T$,
             and  all probability distributions
             are normalized to 1.}
    \be\l{rhodisp}
    \Big\langle \left(\rho-n\right)^2 \Big\rangle =
    2\frac{\langle \Delta_2\{\rho\}\rangle}{
    F_2}~,
    \ee
    where $\Delta_2\{\rho\}$ is given by $\Delta\{\rho\}$, Eq.~(\ref{DF}),
    at the second order,
    \be\l{D2}
    \Delta_2\{\rho\}=
    \frac12 
    F_2 \left(\rho-n\right)^2 
       \ee
(see Ref.~\cite{TR38}, where $A(x)$ is taken here as the free energy
    $F(\rho)$ for a given temperature $T$). 
       Calculating independently the average of $(\rho-n)^2$ in the
       left-hand side (l.h.s.) of
    Eq.~(\ref{rhodisp}) by using approximately
    the probability distribution
    $W_2$ of the second order,
    Eq.~(\ref{W2}), for the particle number density dispersion
    $\mathcal{D}^{(2)}_\rho$ one has
    \be\l{av2}
    \mathcal{D}^{(2)}_\rho\equiv\langle(\rho-n)^2\rangle=
    \int_0^\infty (\rho-n)^2 W_2(\rho)\mbox{d}\rho~;
    \ee
   see Eq.~(\ref{normcond}) for $m=2$.
           Calculating analytically integral in Eq.~(\ref{av2}) at the
   second order
    [Eq.~(\ref{W2})] and comparing the result with  the
    expression on right of Eq.~(\ref{rhodisp}), one obtains (see
    Ref.~\cite{TR38})
    \be\l{difF}
   \langle\Delta\{\rho\}\rangle\approx T/2~. 
    \ee
    For the expressions of the derivatives of the
        free energy $F(\rho)$
    in terms of the incompressibility $\mathcal{K}$ and its second
derivative, one can use the well-known [9] relationship between the
pressure $P(\rho)$ and free energy $F(\rho)$,
\be\l{p}
P(\rho)=-\frac{\partial F}{\partial V},
\ee
where
\be\l{p1}
-\frac{\partial F}{\partial V}=
\frac{\rho^2}{\langle N \rangle} \frac{\partial F(\rho)}{\partial \rho}~.
\ee
   Differentiating the identity (\ref{p}) over $\rho$ and accounting
   also for the zero first derivative
of $F$ at the statistical equilibrium $\rho=n$ in expansion
(\ref{Fexp}), one can express
the second derivative
of $F(\rho)$ over $\rho$ at $\rho=n$ in terms of the
incompressibility $\mathcal{K}$,
\be\l{dFK}
\left(\frac{\partial^2 F(\rho)}{\partial \rho^2}\right)_{\rho=n}=
\frac{\langle N \rangle\mathcal{K}(n)}{n^2} 
\ee
with
\be\l{Kn}
\mathcal{K}(n)=
\left(\frac{\partial P(\rho)}{\partial \rho}\right)_{\rho=n}~.
\ee
Notice that the pressure $P(\rho)$ is an intensive quantity which
depends on particle number $N$ or volume $V$ only through
the particle number density
$\rho$ in our 
    system.
Using Eqs.~(\ref{difF}) and (\ref{dFK}),
from the particle number density dispersion $\mathcal{D}_\rho$,
normalized by $n^2$, Eq.~(\ref{FL-MM}), at the second-order expansion
of the free energy, $\mathcal{D}_\rho^{(2)}$, Eq.~(\ref{D2}),
and 
Eq.~(\ref{W2}) for the probability distribution $W_2$, one obtains
\be\l{omrho2}
\frac{\mathcal{D}_\rho^{(2)}}{n^2}=\frac{T}{\langle N\rangle\mathcal{K}}~.
\ee
Transferring this expression into the particle number dispersion
$D_N$, Eq.~(\ref{FL-gendef}), normalized by $\langle N\rangle^2$, for an
intensive
system, one has
\be\l{D2DN}
\frac{\mathcal{D}_\rho}{n^2}\approx \frac{\mbox{D}_N}{\langle N\rangle^2}.
\ee
Using finally the normalization of the dispersion $\mbox{D}_N$
by $\langle N\rangle$
we arrive at the well-known 
\cite{TR38,RJ58,LLv5,TK66,IA71,BR75,AC90,ZM02} 
 local 
expression
for the fluctuations $\omega$ in terms of the
local isothermal incompressibility,
$\mathcal{K}(T,\rho)$: 
\be\l{FL-press}
\omega(T,n) \approx
\frac{\mbox{D}_N}{\langle N\rangle} \approx \frac{T}{\mathcal{K}}~,\quad
  \mathcal{K} = \left(\frac{\delta P }{\delta n }\right)_T~,
\ee
where $P$ is the pressure, $P(T,n)$, Eq.~(\ref{p}) , and
$P=P(T,n)$ is the equation of state in canonical variables.

Notice that this quadratic-approach result
coincides with the well-known
traditional result of the Landau theory of classical fluctuations \cite{LLv5}.
Landau \cite{LLv5} 
normalized particle number dispersion, $\mbox{D}_N$  by $\langle N\rangle$.
However, the result for the fluctuation $\omega $, Eq,~(\ref{FL-press}),
                $\propto 1/\mathcal{K}$, is 
                divergent near the critical point where $\mathcal{K}=0$.
                In order to improve these fluctuation results we will expand
                the free energy $F$ up to high-order terms in
                Eq.~(\ref{Fexp}) (Sec.~\ref{sec-5}).
                But first, in the next section, let us study the traditional
                result (\ref{FL-press}) in more details.

  \section{Particle number fluctuations and incompressibility}
  \label{sec-4}

  As shown in Sec.~\ref{sec-3} and Appendices \ref{appA} and \ref{appB},
  the fluctuations of particle numbers, $\omega$
  [see Eq.~(\ref{FL-press})],
can be expressed in terms of the
isothermal incompressibility
   $\mathcal{K}$.
We will
  compare the results obtained by employing different approximations
   near the critical point.
Note that in 
the derivations of both formulas, Eqs.~(\ref{FL-sus}) and (\ref{FL-press}),
any fluctuations
  are assumed to be small.
   Nevertheless, we will 
  compare the results obtained by employing different approximations
  near the critical point with the popular formulas given by
  Eqs.~(\ref{FL-sus}) and (\ref{FL-press}), though the value of $\omega$
  near the CP obtained by these formulas is expected to be large.
    In fact, it is still an open
   question whether the
 traditional formula (\ref{FL-press}) can be applied near
 the critical point in
the density-temperature ($n-T$) plane?

For calculations of the incompressibility  $\mathcal{K}$,
one can use 
   the so-called \cite{marik} quantum van der Waals (QvdW)  or
\cite{satarov}
quantum Skyrme local density (QSLD) interaction approaches; see equations
of state (\ref{Pvdw-n}) or (\ref{PQSkyr}), respectively. 
    In this section,
we will follow
Refs.~\cite{FMG19,FMG22} at 
  first order of the
  quantum statistics expansion (see Appendix \ref{appC}).  
 Moreover, this analytical approach will be applied to
 the simplest
 uniform one-component  intensive
 system of
 nucleons
    interacting through the repulsive and
    attractive effective forces.
  For this purpose, the 
     incompressibility
     $\mathcal{K}$ 
     will be considered as 
   a linear and (slightly) nonlinear  response
  of the pressure $P$ to the particle
  number density $n$ variations for the
  nucleon system at constant temperature $T$.
  Our purpose in this section is
  to 
   derive analytical results for the
  fluctuations $\omega$ of particle
      numbers 
      near
  the critical point (CP)
    within the QvdW and QSLD approaches
    at first order of
    the quantum statistics parameter; 
    see
    Ref.~\cite{FMG22}.

For relatively small fluctuations  of 
particle numbers, $\omega$,
one  has Eq.~(\ref{FL-press}).
 In Eq.~(\ref{FL-press}),
$P$ is the pressure for the equation of state, which is given in the
one-component
QvdW and QSLD models
  for symmetric nucleons matter by
  Eqs.~(\ref{Pvdw-n}) and (\ref{PQSkyr}), respectively.
 Notice that we use a more general definition of the
incompressibility $\mathcal{K}$ as the variation derivative of the
pressure of the equation of state over the particle number density $n$
at constant temperature,
in contrast
to its standard definition
as the following first partial derivative of pressure
(see Appendix \ref{appA}).
With this approximation for the incompressibility, $\mathcal{K}_1$,
 one finds
from
Eq.~(\ref{FL-press}) 
the expression for a magnitude of fluctuations:
\be\l{incompexp1}
\omega\approx\omega^{}_1=\frac{T}{\mathcal{K}_1},\quad \mathcal{K}_1 
=\left(\frac{\partial P}{\partial n}\right)_T~.
\ee
 Equation~(\ref{FL-press}) can be first derived from Eq.~(\ref{FL-sus})
in terms of the susceptibility $\chi$ (see Appendix \ref{appA}).
As shown in Appendices \ref{appA} and \ref{appB},
using then linear variations for the
chemical potential $\mu$ as function
of the particle number density $n$,  which are therefore valid
    for small
    fluctuations,  one can
    derive Eq.~(\ref{incompexp1}).
 The value of this fluctuation, $\omega^{}_1$, 
diverges in the CP limit, $n\rightarrow n_c$
and $T\rightarrow T_c$. Therefore, it can be considered only on a finite 
 [sufficiently large for applications of Eq.~(\ref{incompexp1}) but small 
for using expansion near the CP] 
    distance from
the CP. 
The incompressibility $\mathcal{K}$ 
in
Eq.~(\ref{FL-press}) as function of the density $n$ and temperature $T$,
can be expanded in power
series near the critical point $T_c, n_c$ over both variables $T$ and $n$.
The derivatives are evaluated here at the current point $T,n$ within
the precision of high-order terms. 
    Up to second-order terms,
    one has
\bea\l{incompexpfull}
&\omega\approx \omega^{}_3= \frac{T}{\mathcal{K}_3}, \quad \mathcal{K}_3 
=\left(\frac{\partial P}{\partial n}\right)_T +
\left(\frac{\partial^2 P}{\partial n^2}\right)_T (n-n_c)\nonumber\\
&+
  \frac{\partial^2 P}{\partial n\partial T}(T-T_c) +
  \frac12\left(\frac{\partial^3 P}{\partial n^3}\right)_T
  \left(n-n_c\right)^2 ~.
\eea
As suggested in Refs.~\cite{TR38,RJ58}, we use approximately
 the following definition, 
valid at the critical
point\footnote{The CP is assumed to be of the simplest
   second-order,
  in contrast to a high-order
  CP when high-order derivatives become also zero.}:
\be\label{CP-eq}
\left(\frac{\partial P}{\partial n}\right)_T = 0~,~~~~
\left(\frac{\partial^2 P}{\partial n^2}\right)_T=0~.
\ee
 Assuming (see Refs.~\cite{TR38,RJ58}) then that the variations linear in
temperature  and 
quadratic in density are dominant
over high- and low-order variations, one can define
another approximation 
 near the CP:
\bea
&\omega\approx \omega_3^{(c)}=\frac{T}{\mathcal{K}^{(c)}_3}~,\l{om3c}\\
&\mathcal{K}^{(c)}_3=
  \frac{\partial^2 P}{\partial n\partial T}(T-T_c) +
  \frac12\left(\frac{\partial^3 P}{\partial n^3}\right)_T
  \left(n-n_c\right)^2~\l{incompexp}.
\eea
As mentioned above, in these both 
approaches,
Eqs.~(\ref{incompexpfull}) and
(\ref{om3c}) [with Eq.~(\ref{incompexp})], to the variation of the
incompressibility $\mathcal{K}$ and 
of the fluctuations $\omega$, Eq.~(\ref{FL-press}),
all the derivatives are still taken at a current point $n,T$.

\subsection{Fluctuations within the QvdW approach}
\l{subsec-4a}

For the Fermi statistics parameter in the Quantum
van der Waals (QvdW) model, one finds \cite{FMG19,FMG22}
\be\label{del}
\delta  \equiv \frac{\varepsilon}{1-bn}~,\quad
 \varepsilon=\frac{ \hbar^3\,\pi^{3/2}~n}{2\,g\,(mT)^{3/2}}~,
 \ee
  where $m$ is the particle mass and $g$ 
  the system degeneracy.  
Substituting  Eq.~(\ref{Pvdw-n}) for the pressure $P_{\rm W}$ in the
fluctuation $\omega^{}_1$, Eq.~(\ref{incompexp1}),
one can now
obtain $\omega^{}_1$ at the first order over a small
quantum-statistics parameter $\delta$, Eq.~(\ref{del}),
in 
the explicit analytical form
\begin{equation}
  \omega (T,n)\approx \omega^{}_1 =
  \left[(1+2\delta)/(1-nb)^2-2na/T\right]^{-1}~,
 \label{omega2}
\end{equation}
where $\delta$ is given by Eq.~(\ref{del}), $\delta=\delta(T,n)$.
 Then the
behavior of  $\omega(T,n)$, Eq.~(\ref{omega2}), near the critical point
($T_c,n_c$; see 
 Eq.~(\ref{CP-1}) for the first-order analytical CP expressions
    \cite{FMG19,FMG22})
 will be studied within the QvdW model.
 Expanding now the incompressibility $\mathcal{K}$  in powers of the 
     temperature difference $T-T_c$
      [using the expression in the parentheses of Eq.~(\ref{omega2})
     for the fluctuations $\omega^{}_1$],
   we calculate immediately derivatives of the pressure
$P_{\rm W}$, Eq.~(\ref{Pvdw-n}), over the density $n$
 at a current $T,n$ point.
  With the help of the new variables,
\be\l{taunu}
  \tau \equiv ~T/T_c-1~, ~~~~
  \nu \equiv ~n/n_c-1~,
  \ee
    one can fix first
    $n=n_c$ ($\nu=0$) and
        find the behavior  of $\omega(T,n)$ as function of temperature $T$
    near the critical point.
For this purpose, it is convenient to present
$\delta (T,n)$, Eq.~(\ref{del}), as 
\begin{equation}
              \delta (T,n)=\delta \left[(1+\tau) T_c,(1+\nu) n_c\right]~.
   \label{deltataunu}
  \end{equation}
We will take
now the limit of this expression at $\nu=0$ and 
  a small $\tau$.
 In this case, $\nu=0$, 
 one can approximate 
$\delta (T,n)$  at the first-order expansion over $\tau$
 by
  \begin{equation}
   \! \delta ((1+\tau)T_c,n_c) \!\approx\!
    \frac{\hbar^3 \pi^{3/2}n_c}{2g~(mT_c)^{3/ 2}\left(1\!-\!\beta\right)}
    \left(1- \frac{3}{2}\tau \right)~,   
 \label{deltatau}
  \end{equation}
 where  $\beta=b n_c$.

The critical point (CP) of the liquid-gas phase transition satisfies
the 
well-known 
equations (\ref{CP-eq}) 
(see Ref.~\cite{LLv5}).
 Using now Eq.~(\ref{omega2}) and the first of these CP equations   near the CP,
 one finds ($\nu=0$)
\be\label{omegaproxtau}
\omega 
\approx \omega_1^{(c)}((1+\tau) T_c,n_c)=
  \frac{T_c n_c}{P_c } ~ G_{{\rm W},\tau}~ \tau^{-1},
   \ee
where    
  \begin{equation}
G_{{\rm W}\tau} \approx \frac{P_c}
    {T_c n_c}~\frac{\left(1-\beta\right)^2}{1-
        \delta^{}_c}~,\quad  \delta^{}_c=
\delta(T_c,n_c)~.   
 \label{Gtau}
\end{equation}
  Taking the exclusion-volume parameter $b$ from Eq.~(\ref{ab}) and the results
  for $T_c$, $n_c$ and $P_c$ for nucleon
  matter ($g=4, ~m=938$ MeV) from Table
  \ref{table-1},
one finally obtains 
$G_{{\rm W},\tau} \approx 0.29$. 
 This value is only slightly different from 
that of $G_{{\rm W},\tau} \approx 0.26$, obtained in
Ref.~\cite{roma}.
For the case of the classical vdW model ($\delta^{}_c=0$), one  
arrives at the well-known result $G_{{\rm W},\tau}= 1/6$.

Similarly, 
using Eq.~(\ref{deltatau}), for the fluctuations $\omega(T,n) $,
 Eq.~(\ref{omega2}), 
 at the second-order expansion over $\nu$,
for the constant $T=T_c$  ($\tau=0$), 
one  finds 
\bea\l{deltanu}
 &\delta (T_c,(1+\nu)n_c) ~\approx
    ~\frac{\hbar^3 \pi^{3/2}\,n_c}{2g~(mT_c)^{3/ 2}\left(1-\beta\right)}~\nonumber\\ 
  &\times \left(1+ \frac{\nu}{1-\beta}+
        \frac{\nu^2~\beta}{(1-\beta)^2} \!\right). 
 \eea
 Finally, using the fluctuations $\omega^{}_1$ [Eq.~(\ref{omega2})]
       and both CP equations of Eq.~(\ref{CP-eq}) near the CP,
one arrives for $\tau=0$ at
\begin{equation}\l{omegaproxnu}
  \!\omega 
  \approx \omega_1^{(c)}(T_c,(1+\nu)n_c)=\frac{T_c n_c}{P_c }~ G_{{\rm W},\nu}
    ~\nu^{-2}~,
\end{equation}
where,
\be\l{Gnu}
G_{{\rm W},\nu}\approx ~\frac{P_c}{T_c n_c }~~
 \frac{\left(1-\beta\right)^4}{3\beta~[
      2\delta^{}_c\left(1+\beta\right)
            +\beta]} \approx 0.33~. 
 \ee
 The last number was obtained by using Eq.~(\ref{ab}) and
 Table \ref{table-1}. 
 For the case of the classical vdW approach ($\delta^{}_c=0$),
 one has
 from
 Eq.~(\ref{Gnu}) that 
 $G_{{\rm W},\nu}= 2/9$, which is the same as that shown in 
 Ref.~\cite{marik}.
\vspace{0.3cm}
\begin{table}[pt]
\begin{center}
\begin{tabular}{|c|c|c|c|c|c|c|}
\hline
Critical point
& vdW
& first-order & numerical  \\
 parameter & Eq.~(\ref{CP-0})
& Eq.~(\ref{CP-1}) & full QvdW  \\
\hline
$T_c$~(MeV) & ~29.2~& ~19.0~& ~19.7~\\
\hline
$n_c$~(fm$^{-3}$) &0.100 & 0.065 & 0.079\\
\hline
$P_c$~(MeV$\cdot$ fm$^{-3}$) & 1.09  & 0.48 & 0.56\\
\hline
\end{tabular}
\vspace{0.2cm}
\caption{{\small
    Results for the CP parameters of the 
    vdW model
    [Eq.~(\ref{CP-0})]
    (second column) and of the QvdW at first order over the
    quantum statistics parameter $\delta$
    [Eqs.~(\ref{CP-1}) and (\ref{del})] (third column) for symmetric nuclear
    matter; see Eq.~(\ref{ab}) for vdW parameters.
    Numerical results
obtained within the  
 accurate QvdW model in Ref.~\cite{marik} 
   are shown in the fourth column.
}}
\label{table-1}
\end{center}
\end{table}

\subsection{Fluctuations within the QSLD approach}
\l{subsec-4b}

Substituting the pressure $P_{\rm Sk}$, Eq.~(\ref{PQSkyr}), onto
Eq.~(\ref{incompexp1}) for the QSLD
fluctuation $\omega^{}_{{\rm Sk},1}$
at the first order over a small
quantum-statistics parameter $\varepsilon$ [see  Eq.~(\ref{del})],   
one obtains $\omega^{}_{{\rm Sk},1}$, also in 
explicit analytical form, 
\begin{eqnarray}
 & \omega^{}_{\rm Sk}(T,n)\approx \omega^{(1)}_{{\rm Sk},1} 
  \nonumber\\
 &\hspace{-0.4cm}=\left[1+2 \varepsilon -
    2\frac{a^{}_{\rm Sk}n}{T}+b^{}_{\rm Sk}(\gamma+1)(\gamma+2)
    \frac{n^{\gamma+1}}{T}\right]^{-1},
 \label{omega2sk}
\end{eqnarray}
where $\varepsilon=\varepsilon(T,n)$ is the quantum-statistics
parameter [see Eq.~(\ref{del})].
Other QSLD interaction parameters, $a^{}_{\rm Sk}$, $b^{}_{\rm Sk}$, and $\gamma$,
are defined in 
  Eq.~(\ref{Skyrpar}). 
For the classical (zero) SLD approximation to 
the fluctuations,
$\omega^{(0)}_{\rm Sk}$,
 one finds
from Eq.~(\ref{omega2sk}) 
\begin{eqnarray}
&  \omega^{(0)}_{\rm Sk}(T,n)\approx \omega^{(0)}_{{\rm Sk},1}
 \nonumber\\
&\hspace{-0.4cm}=\left\{1- 2\frac{a^{}_{\rm Sk}n^{(0)}_{\rm Sk}}{T^{(0)}_{\rm Sk}}+
  b^{}_{\rm Sk}(\gamma+1)(\gamma+2)
  \frac{[n^{(0)}_{\rm Sk}]^{\gamma+1}}{T^{(0)}_{{\rm Sk}}}\right\}^{-1}~.
 \label{omega2sk0}
\end{eqnarray}

As in  Sec.~\ref{subsec-4a},
 the fluctuations $\omega^{}_{\rm Sk}(T,n)$, Eq.~(\ref{omega2sk}),
 near the critical point
$T^{(1)}_{{\rm Sk},c},n^{(1)}_{{\rm Sk},c}$ (see 
 Eq.~(\ref{SkyrCP-1}) for the first-order analytical CP expressions
   \cite{FMG19,FMG22})
 will be derived within the QSLD approach.
 Expanding now the incompressibility $\mathcal{K}$ in terms of powers of the
     temperature difference $T-T_c$
 [see the expression in parentheses of 
     Eq.~(\ref{omega2sk})],
 for the fluctuations $\omega^{}_{{\rm Sk},1}$  with the help of the new
 variables [Eq.~(\ref{taunu})],
    one can, again, fix first $n$ at
    $n=n_c$ ($\nu=0$). In this way, 
    one finds the behavior  of $\omega^{}_{\rm Sk}(T,n)$
    as function of temperature $T$
    near the critical point. The quantum-statistics
    parameter $\varepsilon(T,n)$
        for the QSLD approach
        plays the same role as $\delta(T,n)$ for the QvdW fluctuation
        calculations.
        We have similar
        expressions for them [see Eqs.~(\ref{deltataunu}),
          (\ref{deltatau}), and (\ref{deltanu})]
        where we only need to replace $\delta$ by $\varepsilon$.
 Using now Eq.~(\ref{omega2sk}), and the first equation in
 Eq.~(\ref{CP-eq}), at first order over
 $\tau$ near the CP,
 one 
 finally obtains ($\nu=0$)
\be\label{Skyromegaproxtau}
\omega^{}_{\rm Sk}
\approx
\frac{T_cn_c}{P_c}G_{{\rm Sk},\tau}~ \tau^{-1}~,
   \ee
   with
  \begin{equation}
G_{{\rm Sk},\tau} \approx \frac{P_c}{T_cn_c}\frac{1}{1-
        \varepsilon^{}_c}\approx 0.32 ~,   
 \label{SkyrGtau}
\end{equation}
  where $\varepsilon^{}_c=\varepsilon(T_c,n_c)$; see Eq.~(\ref{del}).
  In these evaluations, we used  the first-order
  critical temperature
  $T^{(1)}_{\rm Sk}$ and particle number density  $n^{(1)}_{\rm Sk}$,
  from Table \ref{table-2}
  for nucleon matter ($g=4$ and $m=938$ MeV), which were obtained in
  Ref.~\cite{FMG22}.
  For the case of the classical SLD model
   ($\varepsilon^{}_c=0$),  one 
arrives at $G_{{\rm Sk},\tau}= 0.27$.

Similarly, 
we use Eq.~(\ref{deltataunu}). 
Replacing then $\delta$
by $\varepsilon$ for the
fluctuations $\omega^{}_{\rm Sk}(T,n) $,
Eq.~(\ref{omega2sk}),  in the second-order expansion over $\nu$
we apply both CP equations
    in Eq.~(\ref{CP-eq}) near the CP
at constant $T=T_c$ ($\tau=0$).
 Finally, for the fluctuations $\omega^{}_{\rm Sk}$
    [Eq.~(\ref{omega2sk})] 
    at $\tau=0$,
 one approximately arrives at 
\be\l{Skyromegaproxnu}
  \omega^{}_{\rm Sk} 
  \approx
  \frac{T_cn_c}{P_c} G_{{\rm Sk},\nu}~\nu^{-2}~,
  \ee
  where
  \be\l{SkyrGnu}
G_{{\rm Sk},\nu}\approx 
 \frac{P_c}{T_cn_c}\frac{2 T_c}{\gamma(\gamma+1)(\gamma+2)b_{\rm Sk}n_c^{\gamma+1}}
 \approx 0.47 ~.
\ee
In the last estimate, we used Eq.~(\ref{Skyrpar}) for the SLD
parameters and Table \ref{table-2} at $\gamma=1/6$.
Notice that the first order
in expansion of the inverse fluctuation $\omega^{-1}$ for $T=T_c$ 
over $\nu$ disappears because of the second equation for
the critical point
in Eq.~(\ref{CP-eq}). 
For the case of the classical SLD model ($\delta^{}_c=0$),
 one finds from
 Eq.~(\ref{SkyrGnu})  
 a slightly different value, $G_{{\rm Sk},\nu}\approx 0.48$.

 As the QvdW and QSLD fluctuations
 $\omega(T,n)$, 
   Eqs.~(\ref{omega2}) and (\ref{omega2sk}) respectively,
  are functions of the two variables $T$ and $n$,
  one needs to introduce the two-dimensional
   critical-order index, 1 and 2.
 The first 
 component 
 is related to the fluctuation change along the $T$ axis and the second one is
 along the $n$ axis of the $T,n$ range.
  Another characteristic of the
 critical point ($T_c,n_c$) in the $T-n$ plane is the two-dimensional
 fluctuation-slope
 coefficient 
 $\{G_{{\rm W},\tau},G_{{\rm W},\nu}\}\approx\{0.29,0.33\}$ for the QvdW and
 $\{G_{{\rm Sk},\tau},G_{{\rm Sk},\nu}\}\approx \{0.32,0.47\}$ for the
  QSLD approach. 
 For the QvdW case, the slopes $\{G_{{\rm W},\tau},G_{{\rm W},\nu}\}$
 depend on the vdW interaction parameters through
 the critical values $T_c$ and $n_c$  (and therefore $P_c$) and explicitly through the
 exclusion-volume constant $b$. In the case of the QSLD approach,
 $\{G_{{\rm Sk},\tau},G_{{\rm Sk},\nu}\}$, one obtains their dependence on the
 interaction parameters ($a^{}_{{\rm Sk}}$, $b^{}_{{\rm Sk}}$,
 and $\gamma$) for $G_{{\rm Sk},\tau}$
 [Eq.~(\ref{SkyrGtau})] only through the critical
 temperature $T_c$ and density $n_c$ while
 for $G_{{\rm Sk},\nu}$
 one finds also the explicit dependence on  $b^{}_{{\rm Sk}}$, and $\gamma$.
   Notice also that the
   temperature, $\propto 1/\tau$ [Eqs.~(\ref{omegaproxtau}) and
     (\ref{Skyromegaproxtau})],
and the density, $\propto 1/\nu^2$ [Eqs.~(\ref{omegaproxnu}) and
  (\ref{Skyromegaproxnu})], fluctuation dependence near the CP
can be seen
also from Eq.~(\ref{incompexp}) 
for $\omega^{(c)}_3$. In the 
derivations of fluctuations $\omega^{(c)}_1$ and $\omega^{(c)}_{{\rm Sk},1}$
[Eqs.~(\ref{omegaproxtau}) and
(\ref{omegaproxnu}) and Eqs.~(\ref{Skyromegaproxtau}) and
  (\ref{Skyromegaproxnu}),  respectively]
and $\omega^{(c)}_3$  [Eq.~(\ref{incompexp})]
for the QvdW model we
    used 
 Eq.~(\ref{CP-eq}) for the CP before the CP limit,
in contrast to the $\omega^{}_1$ [Eqs.~(\ref{omega2}) and (\ref{omega2sk})] and $\omega^{}_3$
[Eq.~(\ref{incompexpfull})] approximations. Thus, similarly, one obtains
    qualitatively
the same properties of the fluctuations near the CP
for the
QSLD approach as for the QvdW model.

\vspace{0.3cm}
\begin{table}[pt]
\begin{center}
\begin{tabular}{|c|c|c|c|c|}
\hline
Critical point  & 0th-order   & 1st-order & numerical& $\gamma$\\
parameters & Eq.~(\ref{SkyrCP-0}) & Eq.~(\ref{SkyrCP-1})  & full QSMF&  \\
\hline
$T_{{\rm Sk},c}$~(MeV) & ~20.1~& ~15.1~
& ~15.3~& 1/6~\\
\hline
$n^{}_{{\rm Sk},c}$~(fm$^{-3}$) &0.060 & 0.047
& 0.048&
\\
\hline
$P_{{\rm Sk},c}$~(MeV$\cdot$ fm$^{-3}$) & 0.325  & 0.194
&~-~&
\\
\hline
\hline
$T_{{\rm Sk},c}$~(MeV) & ~25.9~& ~21.2~
& ~21.3~& 1~\\
\hline
$n^{}_{{\rm Sk},c}$~(fm$^{-3}$) &0.065 & 0.059
& ~0.059~&~
\\
\hline
$P_{{\rm Sk},c}$~(MeV$\cdot$ fm$^{-3}$) & 0.560  & 0.421
&~-~&\\
\hline
\end{tabular}
\vspace{0.2cm}
\caption{{\small
    Results for the CP parameters of 
    symmetric
nuclear matter in the quantum-statistics Skyrme 
local-density (QSLD) model; second, third, and fourth columns
 are given for $\gamma=1/6$  in the three upper lines; and 
 $\gamma=1$ in the three bottom lines as shown in the
 fifth column; see Eq.~(\ref{Skyrpar})
and Ref.~\cite{FMG22}.  Zeroth- and first-order results for the CP values
[Eqs.~(\ref{SkyrCP-0}) and (\ref{SkyrCP-1})] are shown in the second
and third columns, respectively.
Numerical results obtained within the 
accurate QSLD model 
    in Ref.~\cite{satarov}
    are shown in the fourth column.
}}
\label{table-2}
\end{center}
\end{table}
%

\section{Improved calculations of the fluctuations 
 near the critical point
  }
\l{sec-5}

As mentioned in 
Introduction, 
we need to improve the calculations of the particle
number
    (density) fluctuations
to avoid their divergence near the critical point having zero
incompressibility, $\mathcal{K}=0$, as seen from Eq.~(\ref{FL-press}).
    In order to clarify the 
behavior of fluctuations near the critical point, one has to expand the free energy
$F(\rho)$ up to high-order terms in expansion (\ref{Fexp}) beyond 
the second-order approach considered in
Sec.~\ref{sec-3}.
These high-order terms are needed
because the second derivative of the free energy
is zero at the CP.
  Assuming that the fourth derivative of
the free energy is not zero at the critical point, one can stop the expansion
(\ref{Fexp}) at the fourth-order term;
see Eqs.~(\ref{DF}) for $\Delta\{\rho\}$ and
(\ref{W4}) for the probability distribution $W_4$. As shown in Appendix
\ref{appD}, for the dispersion $\mathcal{D}_\rho$ at fourth order,
Eq.~(\ref{DF}),
one obtains a more general expression, valid also at the critical point,
\be\l{Dn2gen}
\frac{\mathcal{D}_\rho}{n^2}=\frac{\mathcal{A}}{2n^2}\left(\sqrt{1 +
        \frac{4 \langle \tilde{\Delta}\{\rho\}\rangle}{\alpha}}-1\right)~.
\ee
According to
Eqs.~(\ref{ABdef}) and (\ref{alp}),
$\alpha $ and $\mathcal{A}$ are given by
\be\l{alpABDF}
\alpha=\frac{6\langle N\rangle \mathcal{K}^2}{n^2T\mathcal{K}^{\prime\prime}},
\quad
\mathcal{A}=\frac{12\mathcal{K}}{\mathcal{K}^{\prime\prime}}~.
\ee
 The statistical average of
 the dimensionless free energy difference
 $\tilde{\Delta}\{\rho\}=\Delta\{\rho\}/T$,
 Eq.~(\ref{DF}),
 $\langle \tilde{\Delta}\{\rho\}\rangle$ can be largely
 approximated  within the mean field
 approach by
Eqs.~(\ref{tDelF}), (\ref{tDelF2}), and (\ref{tDelF4}).
We introduced also the dimensionless parameter $\alpha$, Eq.~(\ref{alp}),
which is a measure of the effective distance from the critical point.
Indeed, for large $\alpha$, far from the critical point,
one finds asymptotically,
from a more 
general 
equation (\ref{Dn2gen})
for the dimensionless dispersion $\mathcal{D}_\rho$, the
    following limit:
\be\l{D2lim}
\frac{\mathcal{D}_\rho}{n^2}\rightarrow\frac{\mathcal{D}_\rho^{(2)}}{n^2}=
\frac{T}{\langle N  \rangle~\mathcal{K} },\quad \alpha \gg 1~. 
\ee
The superscript ``m'' in $\mathcal{D}_\rho^{(m)}$ means
    the mth-order term 
    of the expansion of $\mathcal{D}_\rho$, Eq.~(\ref{DF}).
    In particular, for the second-order
    term of this expansion one has $m=2$.
  The particle number dispersion $\mbox{D}_N$, Eq.~(\ref{FL-gendef}),
  can be evaluated from the
  following approximate relationship:
  \be\l{DNDrho}
\frac{\mbox{D}_N}{\langle N\rangle^2}=\frac{\mathcal{D}_\rho}{n^2}~.
  \ee
  Using, then, Eqs.~(\ref{Dn2gen}) and (\ref{DNDrho}), one obtains
  \be\l{DNgen}
\mbox{D}_N \approx \frac{\mathcal{A}\langle N \rangle^2}{2n^2}\left(\sqrt{1 +
  \frac{4 \langle \tilde{\Delta}\{\rho\}\rangle}{\alpha}}-1\right)~.
\ee
For small $\alpha$, near the critical point, 
up to small corrections of high order in powers of $1/\langle N\rangle$, 
from Eqs.~(\ref{DNDrho}) and (\ref{DNgen}) one approximately
arrives at 
another known limit \cite{RJ58}:
\be\l{D4limCP}
\mbox{D}_N \rightarrow \frac{\mathcal{D}_\rho^{(4)}\langle N \rangle^2}{n^2}
= \langle N \rangle^{3/2}~\sqrt{\frac{6T}{
     n^2 \mathcal{K}^{\prime\prime}}}~,
\quad \alpha \ll 1~.
\ee
The superscript $m=4$ in $\mathcal{D}_\rho^{(4)}$ shows
    the fourth-order term of the same expansion (\ref{DF}). 
With the expressions (\ref{alpABDF})
for $\alpha$ and $\mathcal{A}$ in terms of the incompressibility
$\mathcal{K}$ and its
second derivative $\mathcal{K}^{\prime\prime}$, one can rewrite
Eq.~(\ref{Dn2gen}) for the particle number
dispersion in a more explicit way. Taking, for instance, the normalization
of the particle number dispersion $\mbox{D}_N$ [see Eq.~(\ref{DNgen})]
as in Eq.~(\ref{FL-press}), for
the sake of comparison, one obtains
\bea
&\omega=\mbox{D}_N/\langle N \rangle
\l{omgenN}\\
& \approx \frac{6\langle N \rangle\mathcal{K}}{n^2\mathcal{K}^{\prime\prime}}\left(\sqrt{1 +
  \frac{2 \langle \tilde{\Delta}\{\rho\}\rangle n^2 T\mathcal{K}^{\prime\prime} }{
    3\langle N \rangle \
\mathcal{K}^2}}-1\right)~\l{Dn2genK};
\eea
see 
Eqs.~(\ref{tDelF})-(\ref{tDelF4}) for expressions for
    $\langle\tilde{\Delta}_F\rangle$ as
functions of $\alpha$.
According to Eqs.~(\ref{D2lim}) and (\ref{D4limCP}), one finds the limits
to the expressions for the two asymptotical
traditional ($\alpha \gg 1$) and improved ($\alpha \ll 1$) dispersions
in a more explicit form:
\bea
&\omega=\frac{\mbox{D}_N}{\langle N\rangle}
\rightarrow\frac{\langle N \rangle\mathcal{D}_\rho^{(2)}}{n^2}=
\frac{T}{\mathcal{K} },\quad \alpha \gg 1~, \l{lim2}\\
&\omega = \frac{\mbox{D}_N}{\langle N \rangle}
\rightarrow\frac{\langle N \rangle\mathcal{D}_\rho^{(4)}}{n^2}=
\sqrt{\frac{6\langle N \rangle T}{
     n^2 \mathcal{K}^{\prime\prime}}}~,
\quad \alpha \ll 1~.\l{lim4}
\eea
Both these limits, far from the critical point ($\alpha \gg 1$)
and near the CP ($\alpha \ll 1$), are well known; see
Refs.~\cite{LLv5} and \cite{RJ58} (Appendix ~\ref{appE}),
respectively, also
Ref.~\cite{TK66} (Appendix~\ref{appF}).
The limit for $\alpha \ll 1$ in Eq.~(\ref{lim4}) to the CP
is the same as in Ref.~\cite{RJ58} if we neglect the second-order term
and keep only the fourth-order
 component in derivations
of Sec.~\ref{sec-3} and Appendix~\ref{appD}; see more
details in Appendix~\ref{appE}.
    \begin{figure}
  \begin{center}
     \includegraphics[width=8.5cm,clip]{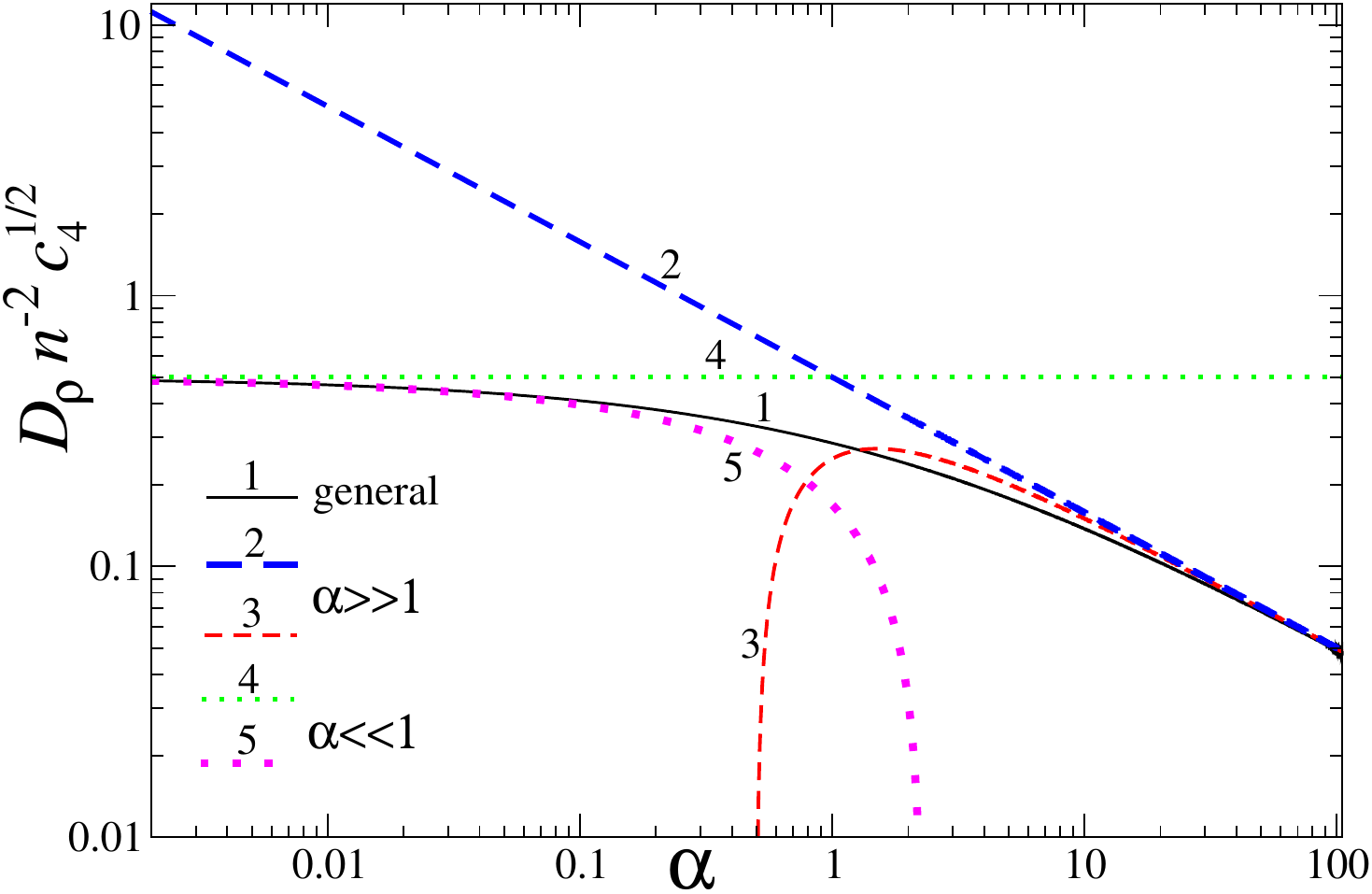}   
    \end{center}
  \caption{ The particle number density dispersion, $\mathcal{D}_\rho$,
    normalized by the factor $n^2/\sqrt{c^{}_4}$, Eq.~(\ref{cm})
    with (\ref{Fm}),
    as a function of the parameter $\alpha$, Eq.~(\ref{alp})
    [in Eq.~(\ref{alpABDF})],
      for a symmetric nucleon system. Solid line ``1'' shows
    the 
   generalized formula (\ref{Dn2gen}). Dashed lines (rare ``2'' and
    frequent ``3'')
    show asymptotes (the main term and that with the
    first correction) in expansion over
    $1/\alpha$) at
    $\alpha \gg 1$, valid far from the critical point, respectively;
    see Eq.~(\ref{Dasinf}).
    Dotted lines (frequent ``4'' and  rare ``5'') present the
    opposite asymptotes
    (the main constant term, 1/2, and that with the
    first correction in expansion
    over $\sqrt{\alpha}$), Eq.~(\ref{Das0}), respectively.
    }
\label{fig1}
\end{figure}
    %
    \begin{figure}
  \begin{center}
    \includegraphics[width=8.5cm,clip]{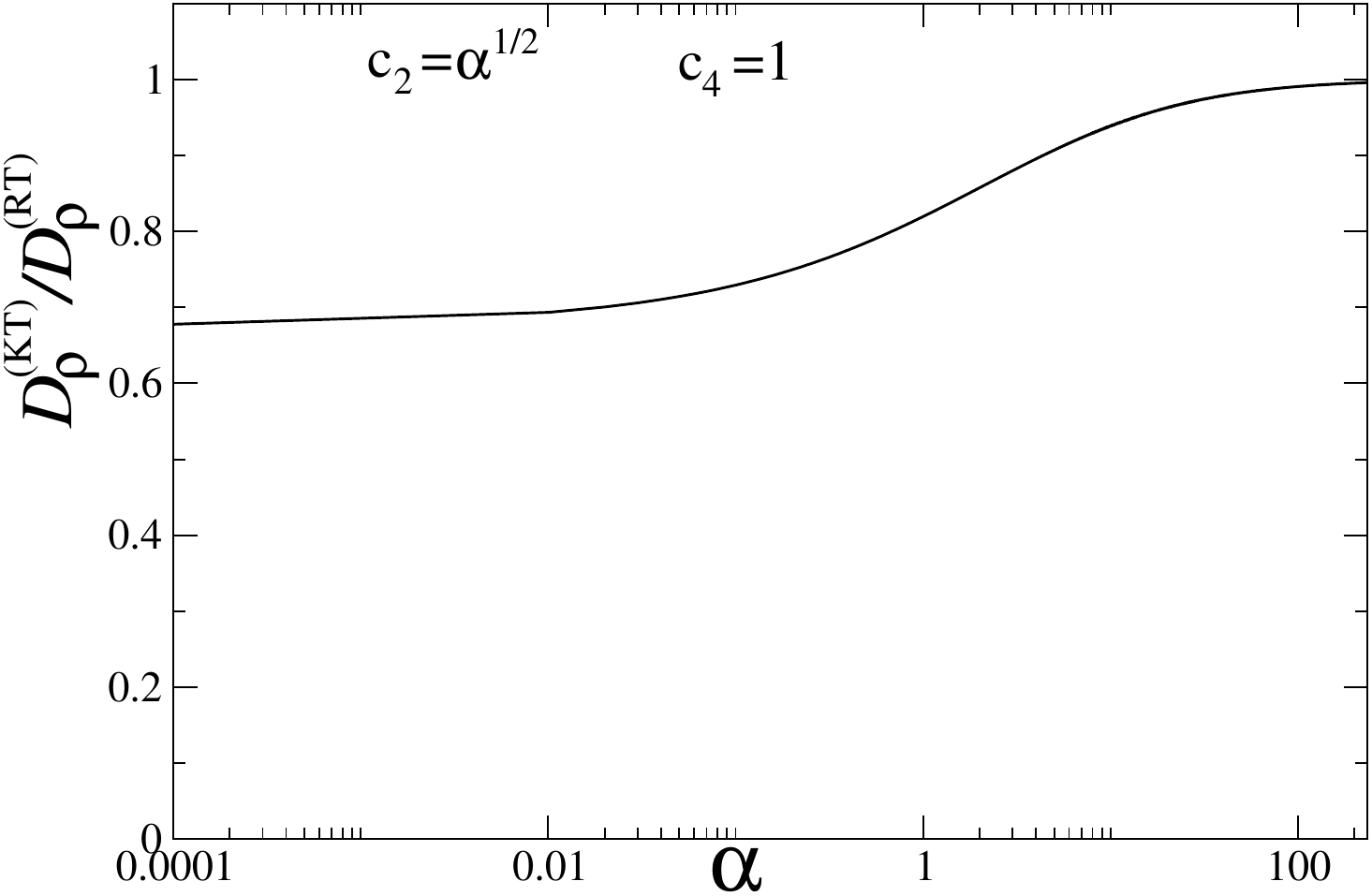}
    \end{center}
  \caption{ Ratio of the particle number density dispersion,
    $\mathcal{D}^{(KT)}_\rho$, Eq.~(\ref{Drhom}) with Eq.~(\ref{m}),
    in the K.~Tolpygo (KT) approach
    (Appendix~\ref{appF})
    to that of $\mathcal{D}^{(RT)}_\rho$, Eq.~(\ref{Dn2gen}),
    in the R.~Tolman (RT) approach
    (Appendix~\ref{appD} and Fig.~\ref{fig1})
    as a function of the same parameter $\alpha$, 
    Eq.~(\ref{alp}),  or 
      Eq.~(\ref{alpha}) for $c^{}_4=1$  and $c^{}_2=\sqrt{\alpha}$ for the
      same system as that of Fig.~\ref{fig1}.
      }
\label{fig2}
\end{figure}

    Figure~\ref{fig1} shows the dependences of the
    generalized expression (\ref{Dn2gen})
(solid line ``1'') for the
 dispersion $\mathcal{D}_\rho$ 
in the units, explained in the caption,
as a function of the critical parameter $\alpha$.
The dashed (``2'' and ``3'') and dotted (``4'' and ``5'') lines present
the asymptotes for $\alpha \gg 1$ and $\alpha \ll 1$, for the main term
and its first correction, respectively:
\bea\l{Das}
&\mathcal{D}_\rho
\rightarrow \frac{n^2}{c^{1/2}_4}\frac{1-1/(2\alpha)}{2\sqrt{\alpha}},~~~
\alpha \gg 1~,\l{Dasinf}\\
&\mathcal{D}_\rho \rightarrow
\frac{n^2}{c^{1/2}_4}\left(1/2-q \sqrt{\alpha}\right)~,~~~\alpha \ll 1~.
\l{Das0}
\eea
where
\be\l{q}
q=2\frac{[\Gamma(1/4)+\Gamma(3/4)]\Gamma(5/4)}{4 \Gamma(1/4)\Gamma(5/4)}
    \approx 0.331~,
\ee
and $\Gamma(x)$ is the standard gamma function.
As seen from this figure, lines ``2'' and
``5'' show the well-known asymptotic 
results, Eqs.~(\ref{D2lim}) and
    (\ref{D4limCP}), in Refs.~\cite{TR38,RJ58}.
    The convergence is seen even better if we take into
    account also the first corrections
    to these main components of the asymptotes
    [see
    Eqs.~(\ref{Das}) and (\ref{Das0})].  
    We formally prolonged them analytically to other
    values of $\alpha$ far away from the limit boundaries,
    where we have  
   main asymptotes [Eqs.~(\ref{D2lim}) and
    (\ref{D4limCP})]. The reason is to find the values of $\alpha $ where
    one finds 
    their convergence to a more general formula (\ref{Dn2gen})
    far from ($\alpha \gg 1$) and
    near ($\alpha \ll 1$) the critical point, relatively.
    As seen from this figure, one can see their convergence
    at the ends of the shown interval of $\alpha$. In some sense, the formula
    (\ref{Dn2gen}), derived in Appendix~\ref{appD} in the mean field
    approach \cite{huang} neglecting density-density correlations, is
    more ``universal'' than the traditional fluctuation
    formula (\ref{FL-press}), as
    an analytical transition of the result for
    fluctuations $\omega$ from that at the
    critical point ($\alpha=0$) to the result of Eq.~(\ref{FL-press}).
    The tradition formula (\ref{FL-press}) is
    valid in fact far from the critical point
    at $\alpha \gg 1$ at any large but finite particle number average
    $\langle N \rangle$.  We emphasize that this transition is
    presented independently of
    the specific effective interactions. Thus, 
    from this figure, one can evaluate the values
    of $\alpha$, as a measure of the distance from the critical point,
    for which one can use the asymptotes (\ref{lim2}) and (\ref{lim4}).
\noindent
 The expressions (\ref{Dn2genK}) 
and, in particular, (\ref{lim4}) can be finite at the critical point if the
second derivative of the incompressibility $\mathcal{K}$,
$\mathcal{K}^{\prime\prime}$,
is not zero. As noticed 
in Refs.~\cite{RJ58,TK66}, the result,
Eq.~(\ref{lim4}), for some intensive systems
agrees better with the experimental data on
    opalescence 
    than the traditional
    Eq.~(\ref{FL-press}).
    However, we should note that the approaches used in
    Refs.~\cite{TR38,RJ58} (Appendices~\ref{appD} and \ref{appE})
    and in Ref.~\cite{TK66} (Appendix~\ref{appF})
    with approximately the same probability distribution $W_4$,
    Eq.~(\ref{W4}), for calculations of the statistical averaged
    dispersion,
    or variance $\langle (\rho-n)^2\rangle$, are somewhat different.
    As shown in Appendix~\ref{appD} for the derivations by the Tolman
    approach \cite{TR38}, the statistical consistency condition (\ref{conseqx2})
    for the quantity $(\rho-n)^2$ in average, $\langle (\rho-n)^2\rangle$,
    and the mean field approach neglecting density-density 
    correlations, with the expansion
    (\ref{Fexp}), is essentially used, in contrast
    to the Tolpygo approach \cite{TK66}; see Appendix~\ref{appF}.
    This
    explains a difference in analytical results for the dispersion $D_\rho$
    in Appendices~\ref{appD} and \ref{appF}.
    Therefore, the limits of the variance $D_\rho$ in both compared
    approaches,
      to the critical point ($\alpha \rightarrow 0$),
   are different by a constant;
    cf.~Eqs.~(\ref{D4limCP}) and (\ref{D0c2}) (see Fig.~\ref{fig2}).
    As seen from Fig.~\ref{fig2}, the limit for large $\alpha$
    is the same for these compared approaches.
    Another peculiarity of
    our approach to the classical fluctuation theory in
    both discussed versions
    (see Appendices \ref{appD} and \ref{appF})
    is the basic
    one-parameter analytical transition over the effective 
      distance  $\alpha$
    from
     the CP, in contrast to the two-parameter analytical transition
     over $c^{}_2\propto F_2$ and $c^{}_4\propto F_4$, separately.
      Both these approaches are remarkable in showing
     that for the mean field approach (up to the correlations above a
     mean field) for
     the finite particle number average $\langle N\rangle $ of a 
     nuclear matter piece is finite everywhere including the critical point,
    in contrast to the traditional divergent result, $\omega=T/\mathcal{K}$.
     Notice also that in this way we may consider 
     high-order critical points by taking into account high-order
     terms in the expansion  (\ref{DF}) for the free 
     energy.  
     It is clear also how to extend Appendix \ref{appD} to
       the fluctuation results, 
       accounting for even more important effects of 
     density-density correlations.

         So far we did not need to specify the interactions which are presented
     here only in terms of the pressure of the equation of state
     through the incompressibility and its second derivatives.
     Notice also that, for relatively large temperatures
     $T$ and small mean particle-number
    densities  $n$, the quantum
    statistics parameter $\varepsilon$ [Eq.~(\ref{del})] is small.
    Therefore,  in this
        part of the $T$-$n$ plane,
    in
    contrast to the
    calculations of the critical points, for simplicity one can
    neglect the quantum statistics effects 
        in the pressure for 
   approximate evaluations of the fluctuation $\omega$.
    Indeed, as shown in the previous section, the fluctuations $\omega$ within
    the QvdW and QSLD  models
    do not 
    depend much 
    on these effects.
    Therefore,  we will first consider 
    more accurate calculations, near the CP,
    of the fluctuations $\omega $ 
    in terms of the  same vdW and SLD
    pressures of the corresponding equations of state, 
    neglecting  small
    quantum statistics corrections \cite{LLv5,FMG19,FMG22}.

    Substituting now
    the pressure for the vdW equation of state (\ref{Pvdw-n})
    at $\delta=0$
    into Eq.~(\ref{lim4}), valid near the CP, one obtains
    \be\l{FL-4cpvdW}
    \omega=\frac{\left(1-bn\right)^2\sqrt{\langle N \rangle}}{bn}~~~
    (\alpha \rightarrow 0)~~
    \mbox{vdW} ~.
\ee    
Notice that this result is independent of temperature $T$ and of the attractive
vdW constant $a$ but depends on the product of the particle number density $n$
times the repulsive exclusion-volume interaction constant $b$.
It is not the case for the SLD
interparticle interaction. As expected, the value of $\omega$  at the critical
point (\ref{CP-0})
is finite of the order of
or smaller than $\sqrt{N}$,
for a finite average number of particles, $\langle N \rangle$.
More accurately, this value of the fluctuations, $\omega$,  is 
$4 \sqrt{\langle N\rangle}/3$. Notice that this value
 is a little larger
than
that in Ref.~\cite{TK66}
because of reasons explained above in this section;
see Appendices~\ref{appD} and \ref{appF}.
Substituting the SLD equation of state (\ref{PQSkyr}) at $\varepsilon=0$
 into Eq.~(\ref{lim4}), one arrives at
    \be\l{FL-4cpSLD}
    \omega=\sqrt{\frac{6T\langle N \rangle}{
        \gamma(\gamma+1)(\gamma+2)b n^{\gamma+1}}}
    ~~~(\alpha \rightarrow 0)~~\mbox{SLD} ~.
\ee
As seen from this expression, the SLD fluctuations depend on the temperature
$T$ and interaction constants $b$ and $\gamma$, but are independent of the
attractive interaction constant $a$ as in the vdW case. For the value at
the critical point,
one obtains also finite results of the order of $\sqrt{N}$, namely,
$1.45\sqrt{\langle N \rangle}$ for $\gamma=1/6$ and
$0.44\sqrt{\langle N \rangle}$ for $\gamma=1$
for a given value of the particle number average
$\langle N \rangle$. Thus, in contrast
to the traditional expression (\ref{FL-press}), for the fluctuations $\omega$,
Eqs.~(\ref{FL-4cpvdW}) and (\ref{FL-4cpSLD}), valid in
the limit to the CP, depend  on
the mean particle number $\langle N \rangle$ by a factor which is
proportional
to the value of $\sqrt{\langle N \rangle}$.

\begin{figure}
\begin{center}
  \includegraphics[width=8.5cm,clip]{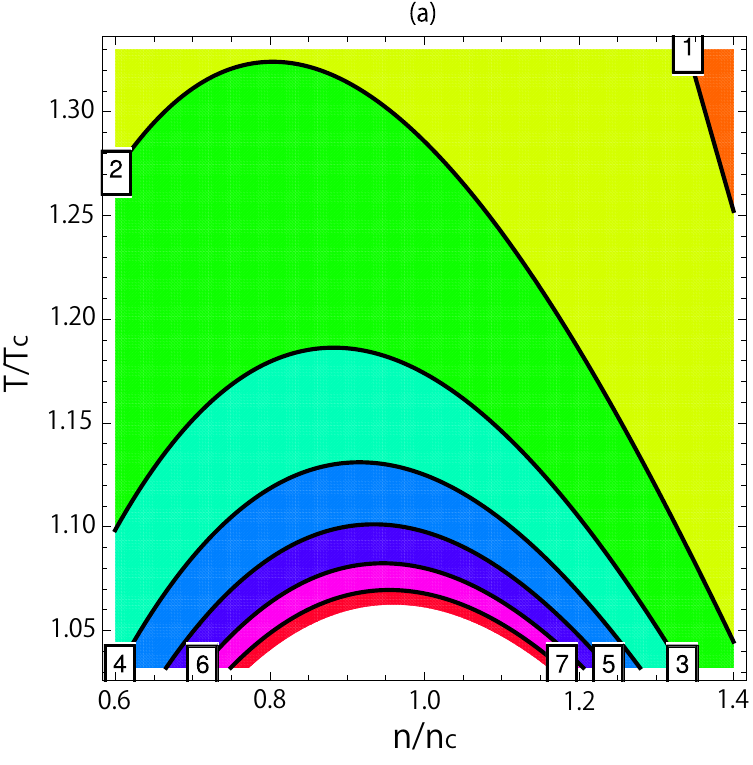}

  \vspace{0.8cm}
  \includegraphics[width=8.5cm,clip]{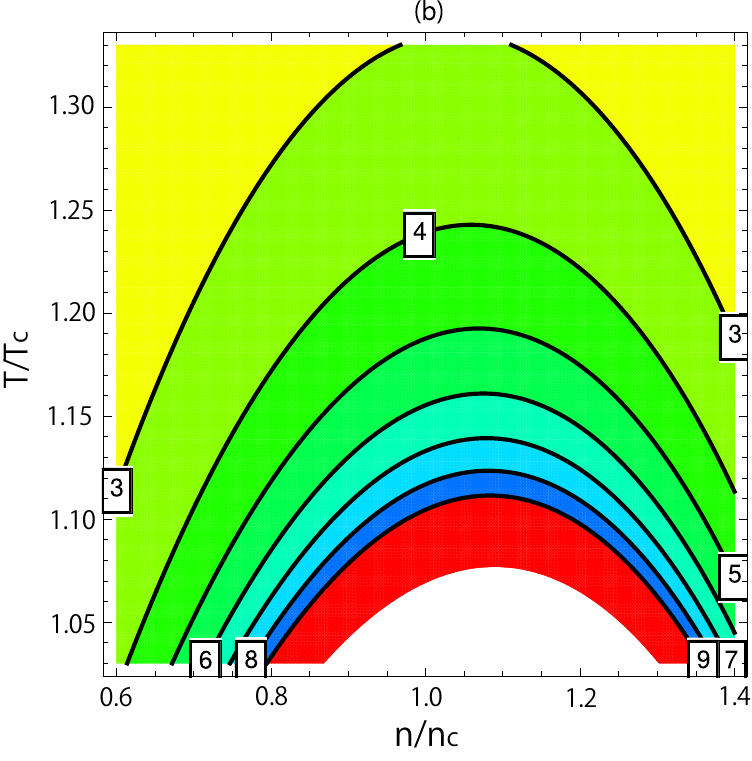}
\end{center}

\vspace{-0.7cm}
\caption{
  Contour plots for the QvdW approximations to the particle number
  fluctuations
  $\omega$ as functions of the averaged
  density $n$ and temperature $ T $ 
  (in units of the corresponding critical values
  $T_c$ and $n_c $) near the critical point.
  The zero approximation, vdW (a), Eq.~(\ref{FL-press}) 
  with the vdW pressure [Eq.~(\ref{vdW})],
  and the first-order QvdW approach (b) in the quantum statistics
  expansion  over a small parameter $\delta$  
  [Eq.~(\ref{omega2})]. 
      Numbers in white squares at lines of constant fluctuations
      $\omega(T,n)$ show their values.
}
\label{fig3}
\end{figure}

\section{Discussion of the results}
\l{sec-6}

 Figure~\ref{fig3} shows the particle number fluctuations
$\omega(T,n)$  as a function of the dimensionless
temperature $T/T_c$ versus
    density $n/n^{}_c$
    variables
    for symmetric nuclear matter 
 by the traditional calculations
     employing 
     Eq.~(\ref{FL-press}). The zeroth-order approximation 
       using Eq.~(\ref{vdW})
[vdW (a)] and the first-order [QvdW (b), Eq.~(\ref{omega2})] approach 
within the quantum statistics
expansion over $\delta $ are shown in these contour plots.
The contour plot of Fig.~\ref{fig3} (b) presents the
calculations of 
fluctuations $\omega^{}_1$ [Eq.~(\ref{omega2})] 
  without using an expansion over a distance from
the critical point. 
As seen from Fig.~\ref{fig3}
[cf. panels (a) and (b)],
the quantum statistics effects in fluctuations $\omega$
are
small, 
as demonstrated by their
numerical values. 
Note that we excluded 
a large shift of the critical point 
by choosing the scaling CP units.
Then, the panels (a) and (b) 
become qualitatively very similar.
 As a function of the density $n$,
the
$\omega^{}_1$ 
contour plot (b)
is approximately
symmetric 
with respect to the CP. The vdW contour plot (b) is only 
 a little asymmetric far from the CP.
As functions of the temperature $T$, 
both plots [(a) and (b)]
are similar but  very asymmetric with respect to $T=T_c$.
Therefore, they are shown only above the
critical point, 
%
\begin{figure*}
  \begin{center}
     \includegraphics[width=8.5cm,clip]{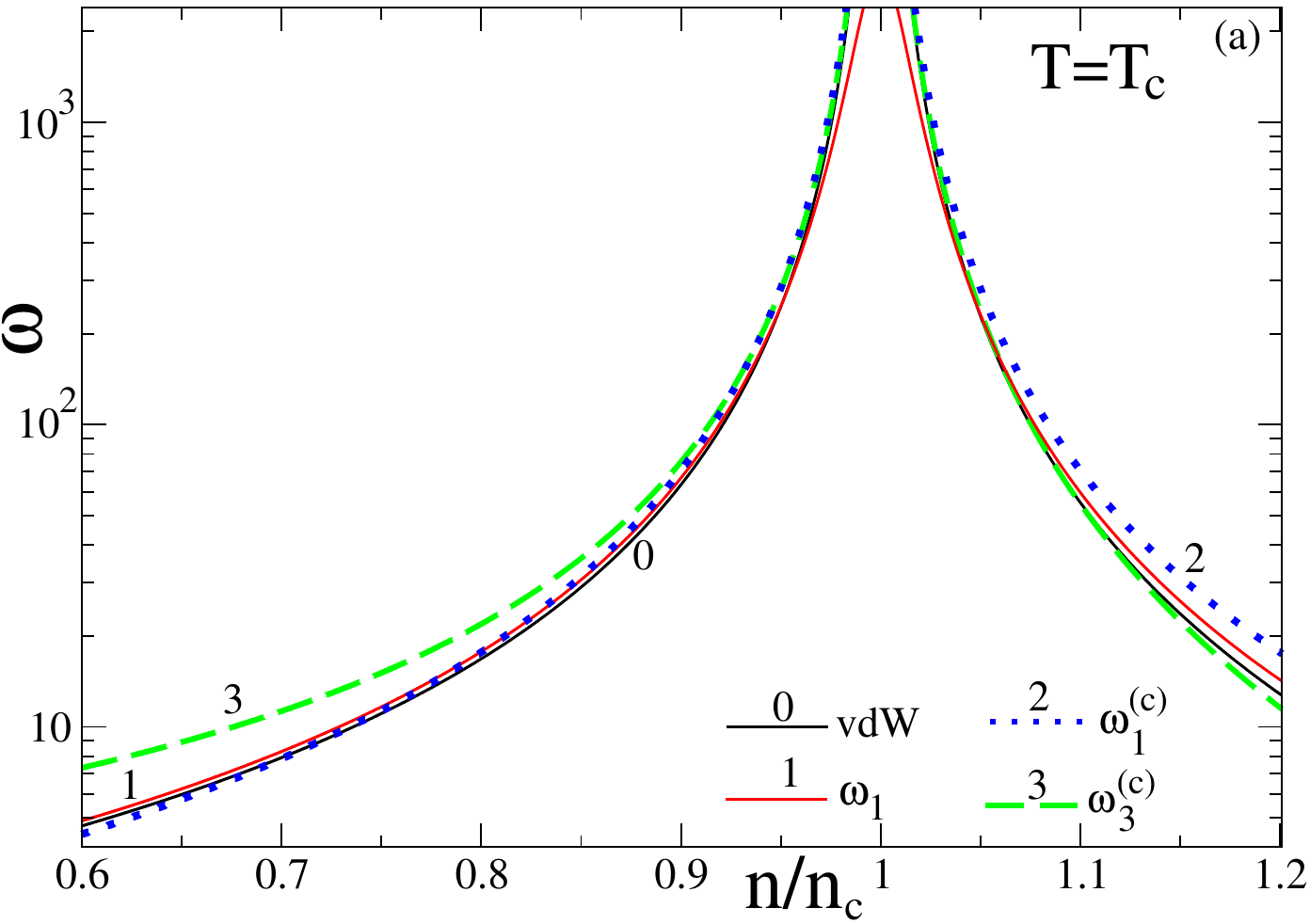}
~    
  \includegraphics[width=8.5cm,clip]{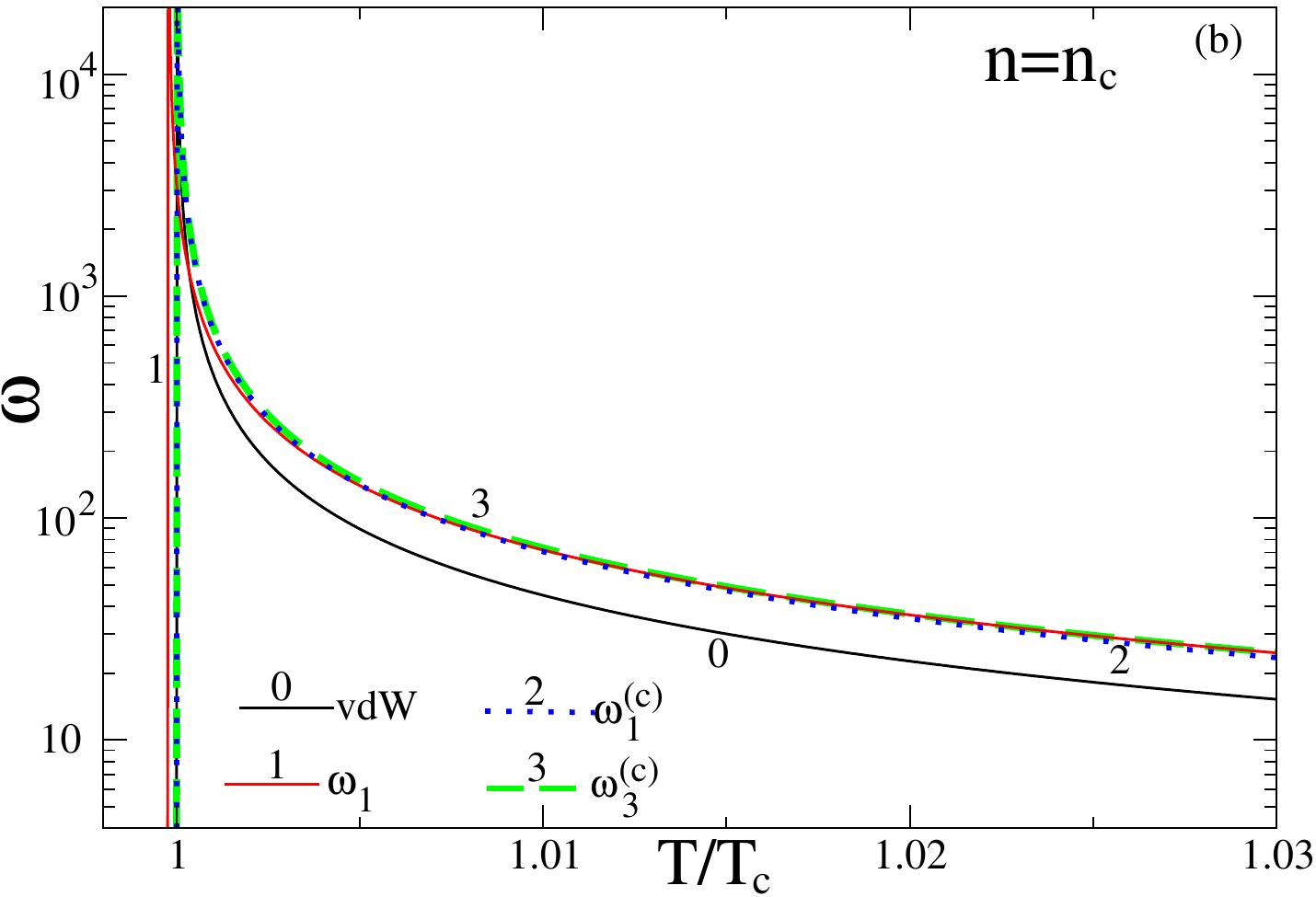}     
\end{center}

\caption{
  Fluctuations of the particle numbers, $\omega$,
  for a nucleon system as
  function of the mean particle-number density $n$ (a) and of
  the temperature $T$ (b) in units of critical values $n_c$ and $T_c$,
  respectively. 
  Solid black lines ``0'' show 
  the zeroth order (vdW), and other lines 
  present different approximations with the first quantum-statistics
  correction; 
  solid red lines ``1'' are the results of calculations by Eq.~(\ref{omega2})
  for $\omega^{}_1$;
  dotted blue
  lines ``2'' show Eqs.~(\ref{omegaproxnu}) in (a) and
  (\ref{omegaproxtau})
  in (b) for $\omega_1^{(c)}$; dashed green lines ``3'' are given by
  Eq.~(\ref{incompexp}) for $\omega^{(c)}_3$.
  }
\label{fig4}
\end{figure*}
\noindent
$T>T_c$. Huge values of the fluctuations near the critical point
    are shown by white regions.
Contour plots for fluctuations $\omega$ at a
few next high orders in the quantum statistics expansion over $\delta$
are visually almost the same as for the first order and,
therefore, 
 are not shown in
Fig.~\ref{fig3}.  

\begin{figure}
  \begin{center}
 \includegraphics[width=8.5cm,clip]{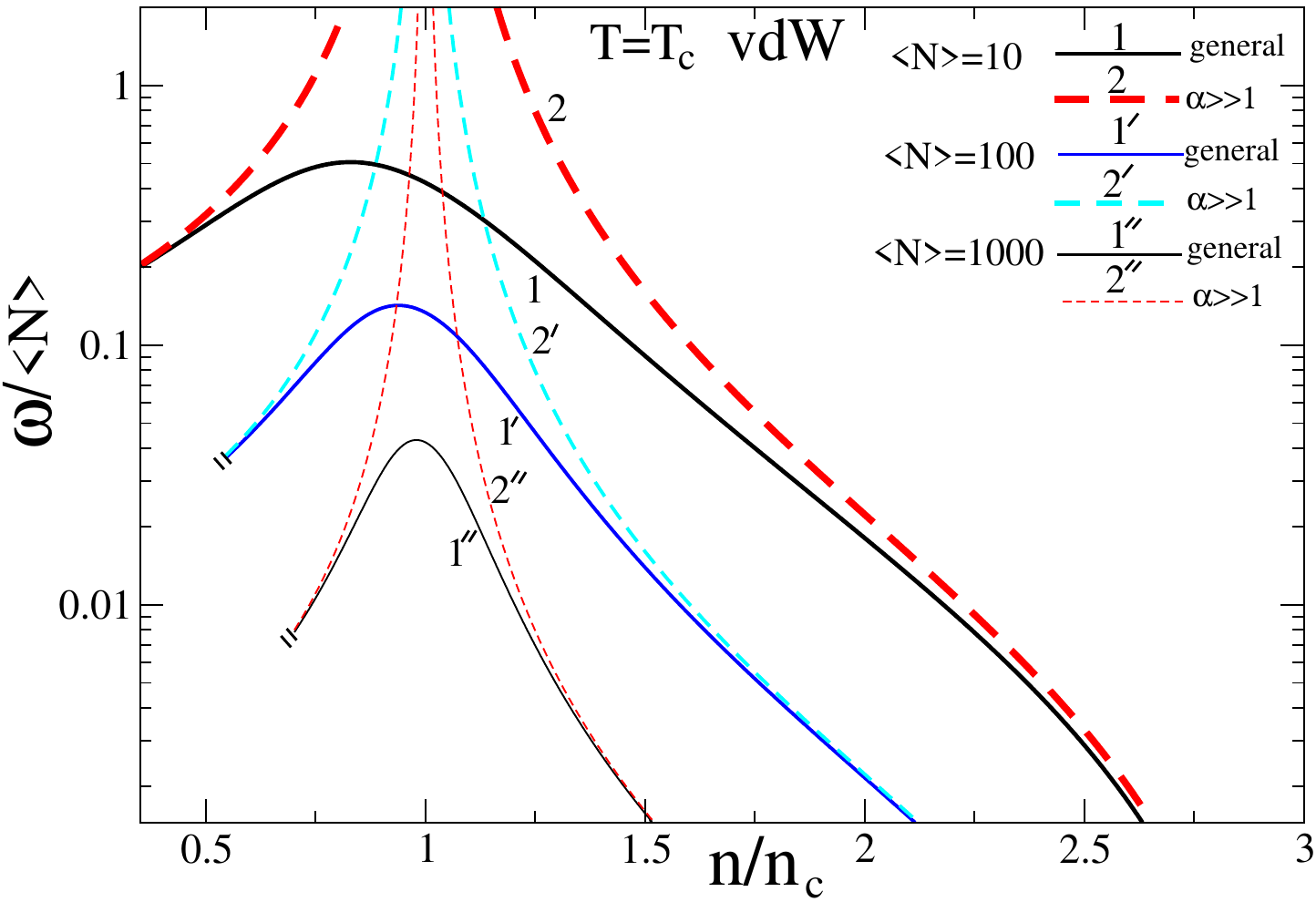}   
\end{center}
  \caption{ 
    The particle number fluctuations 
  $\omega$ 
   [Eq.~(\ref{Dn2genK}), solid lines], divided by constant
    $\langle N\rangle$, i.e., the dispersion $\mbox{D}_N$ 
  normalized by $\langle N\rangle^2$,
are shown as functions of the
average particle number density $n$ (in units of its critical value)
at the critical temperature
$T=T_c$ 
for a symmetric nucleon system with the vdW effective interaction
at different particle number averages $\langle N \rangle$.
Dashed lines present the corresponding
traditional asymptote, Eq.~(\ref{lim2})
  ($\alpha \gg 1$).
     The particle number averages
  $\langle N\rangle =10$ (``1'' and ``2''),
  100 (the same but with primes),
  and 1000  (with double primes) are taken as typical examples.
  Solid lines ``1'', ``1$^{\prime}$'', and ``1$^{\prime\prime}$'' are obtained by
  the generalized formula (\ref{Dn2gen});
  and dashed lines ``2'', ``2$^{\prime}$'',
  and ``2$^{\prime\prime}$'' show the traditional asymptote (\ref{lim2})
  in the same units.
    In order to compare with the traditional approach, 
      the parameters of the vdW effective interactions
  are given by Eq.~(\ref{ab}) [Eq.~(\ref{CP-1}) and Table \ref{table-1}
    for the critical values].
  }
\label{fig5}
\end{figure}
%
\begin{figure}
  \begin{center}
   \includegraphics[width=8.5cm,clip]{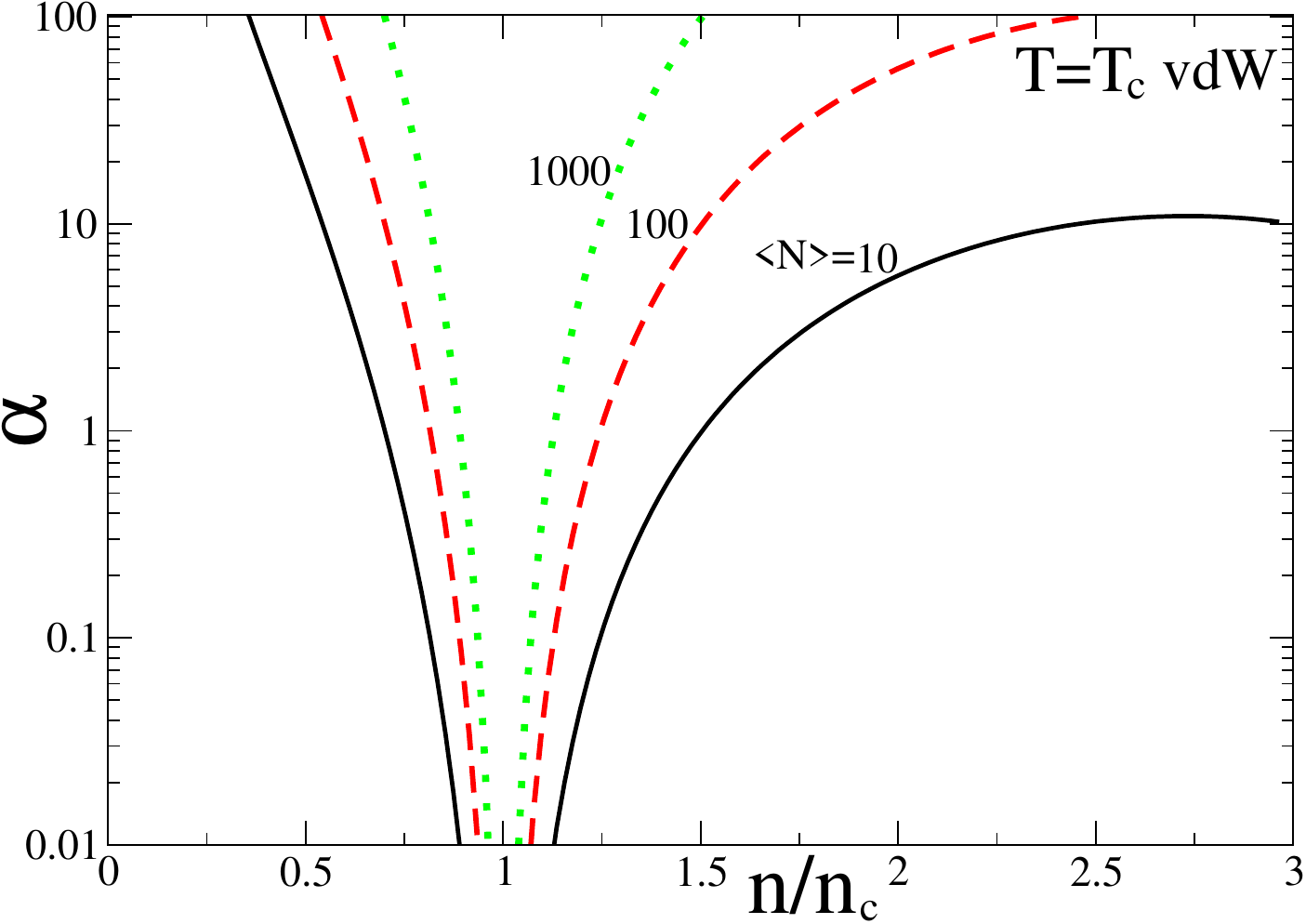}   
\end{center}

  \caption{The parameter $\alpha$, Eq.~(\ref{alp}), as function
    of the particle number average $n$ in the critical value units $n_c$
     at the critical temperature $T=T_c$ for the vdW interaction,          
              and at the same values of the particle number averages
              $\langle N\rangle$.
  }
\label{fig6}
\end{figure}
%
\begin{figure}
  \begin{center}
    \includegraphics[width=8.5cm,clip]{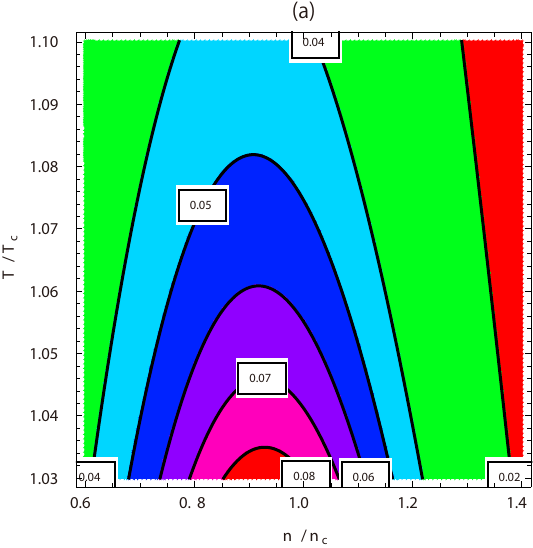}

    \vspace{0.8cm}
     \includegraphics[width=8.5cm,clip]{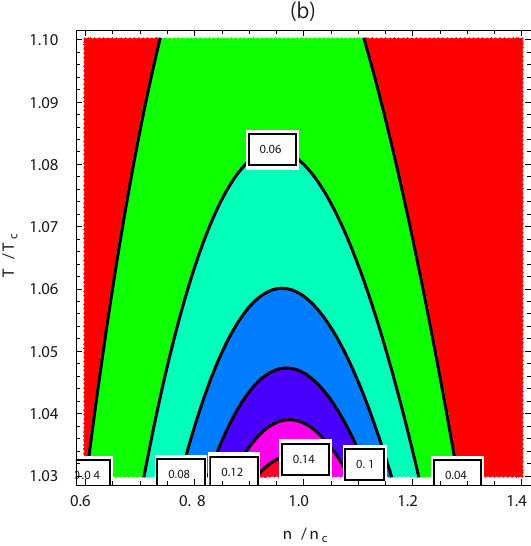} 
\end{center}

  \vspace{-0.7cm}
  \caption{Contour plots for the improved calculations of the fluctuations
    $\omega$ [see Eqs.~(\ref{Dn2genK}) (top) and
      (\ref{lim4}) (bottom)], both divided by the particle number
    average $\langle N \rangle$, as functions
    of particle number density $n$
    and temperature $T$ in their critical values' units. The interval
    of $n/n_c$ is the same as in Fig.~\ref{fig3}. 
    Slightly smaller temperatures $T/T_c$ are taken
    to see more details
    near the critical point. For example, we use $\langle N \rangle=100$
    in these plots.
  }
\label{fig7}
\end{figure}
Figure~\ref{fig4} presents 
a comparison
 between fluctuations $\omega$ using different approximations
  [within Eq.~(\ref{FL-press})]
near the CP, separately,  as functions of 
the mean density $n$, $T=T_c$, panel (a),
and temperature $T$, $n=n_c$,  panel (b), 
both with a better resolution (see Sec.~\ref{subsec-4a}).
 Huge bumps near the CP in the fluctuations $\omega^{}_1$
[solid line ``1'', Eq.~(\ref{omega2})]
 are shown
 in both panels of this figure. Similar bumps appear near the CP
 for fluctuation $\omega^{}_3$ as those in  
 Eq.~(\ref{incompexpfull}) for $\omega^{}_1$, which is not shown therefore in
 Fig.~\ref{fig4}
 for simplicity:
 These two approaches, $\omega^{}_1$ and $\omega^{}_3$, 
 near the
 CP converge to each other in the limit
to the CP with decreasing distance from the CP. 
A divergence of the fluctuations [see Eq.~(\ref{FL-press}) for $\omega$]
at the CP peak is
seen explicitly in 
 the $\omega^{(c)}_1$ ``2'' curves [dashed blue,
Eqs.~(\ref{omegaproxnu}) in Fig.~\ref{fig3}(a)
and
(\ref{omegaproxtau}) in Fig.~\ref{fig3}(b)],
as well as in the $\omega^{(c)}_3$ ``3'' lines' [
  long-dashed green,
Eq.~(\ref{incompexp})]. They
 explicitly diverge at the CP as
in the standard vdW approach (thin black solid line).
Notice that the lines ``2'' and ``3'' converge to each other
better, the smaller the distance is from the CP.
 This is naturally in good agreement with the analytical 
arguments based on Eq.~(\ref{incompexp}), and  in line 
with the arguments 
given in
Sec.~\ref{subsec-4a}.
Such an agreement becomes 
 essentially worse
with increasing distance from the CP.  Both the ``2'' and ``3''
curves have a similar divergent behavior
because in calculations of both curves we neglected first-
and second-order 
derivatives of the isothermal incompressibilities over density $n$
near the critical point, Eq.~(\ref{CP-eq}). 
A huge sharp bump in
the
density ($T=T_c$) (a) and, even much sharper, in the temperature ($n=n_c$)
(b)
dependence for different approximations are
largely in agreement.  
This agrees also
with the accurate numerical
calculations \cite{vova} 
using the same formula (\ref{FL-press}) for the fluctuations $\omega$ at the
incompressibility $\mathcal{K}$, close to zero 
    in the CP limit,
$\mathcal{K}\rightarrow 0$~.
As seen from Fig.~\ref{fig3},
differences between the position of this bump
and CP values for the temperature dependence (b) are more
pronounced in contrast to 
the density function (a).  But, in fact, these differences
are relatively very small within errors
of the derivations
 (see also Fig.~\ref{fig3}).
Note also that the density $n$ behavior (a) is largely symmetric with
respect to the
CP, in contrast to a very asymmetric temperature $T$ dependence (b). This 
 is seen also in the contour plots 
of Fig.~\ref{fig3} where we show $T$
ranges only above the CP.

Notice that it is obviously impossible
to realize practically
the conditions for validity of the considered  approximations to the
fluctuations $\omega $ calculated
in terms of the incompressibility $\mathcal{K}$ 
    by Eq.~(\ref{FL-press})
in the limit to the CP. 
We have to involve more and more terms of 
    expansion of the variation derivative of the incompressibility
    $\mathcal{K}$ [Eq.~(\ref{FL-press})]
    over a distance from the CP.
On the way to the CP,
one has to
stop at small but finite
distance from the CP where a huge bump appears. The 
considered variations
    fail because 
 they become   smaller
or of the order of next derivatives contributions in the expansion
of the incompressibility $\mathcal{K}$ in the denominator of
the fluctuations $\omega$, Eq.~(\ref{FL-press}),
beyond Eq.~(\ref{incompexpfull});
see
Refs.~\cite{TR38,RJ58,TK66}.
 As mentioned above, one may find also arguments 
for validity of
the derivations of Eqs.~(\ref{FL-press}) [or Eq.~(\ref{FL-sus})]
for the fluctuations $\omega$ through the derivatives of
the thermodynamic averages (pressure or particle number density)
 in Refs.~\cite{LLv5,TR38,RJ58,TK66,KG67,ZM02,KW04}.
As emphasized in these works,
large values of relative fluctuations are in
contradiction 
with the basic assumptions
of 
statistical physics because thermodynamic averages, defined up to their
fluctuations,
become meaningless
\cite{LLv5,TR38,RJ58,KG67,ZM02,KW04}.
According to the assumptions in
these derivations 
    (see Secs.~\ref{sec-3} and \ref{sec-4}),
we should have an opposite tendency,  namely, that
    the 
    relative
fluctuations $\omega$ must be 
small, 
in particular near the critical point.
 Therefore, more accurate calculations of the particle
    number fluctuations
        in terms of  
    the statistically averaged Gibbs distribution over particle numbers
     should be considered 
    in a very close range near  the critical point
        of 
        nuclear matter.

    In order to compare with the traditional calculations of
      Sec.~\ref{sec-4} (Figs.~\ref{fig3} and \ref{fig4}),  
    we will discuss now 
   the fluctuations within a more general theory 
             (see Sec. 
              \ref{sec-5}) for the same two simple examples
    of the vdW and SLD
    approaches to the interparticle interactions
    but neglecting small quantum corrections.
We will
discuss then the asymptotic approximations to the generalized formula
(\ref{Dn2gen}) 
for the particle number
fluctuations $\omega$ far from and close to the critical point
    [Eqs.~(\ref{lim2}) and (\ref{lim4})].
%
\begin{figure}
  \begin{center}
   \includegraphics[width=8.5cm,clip]{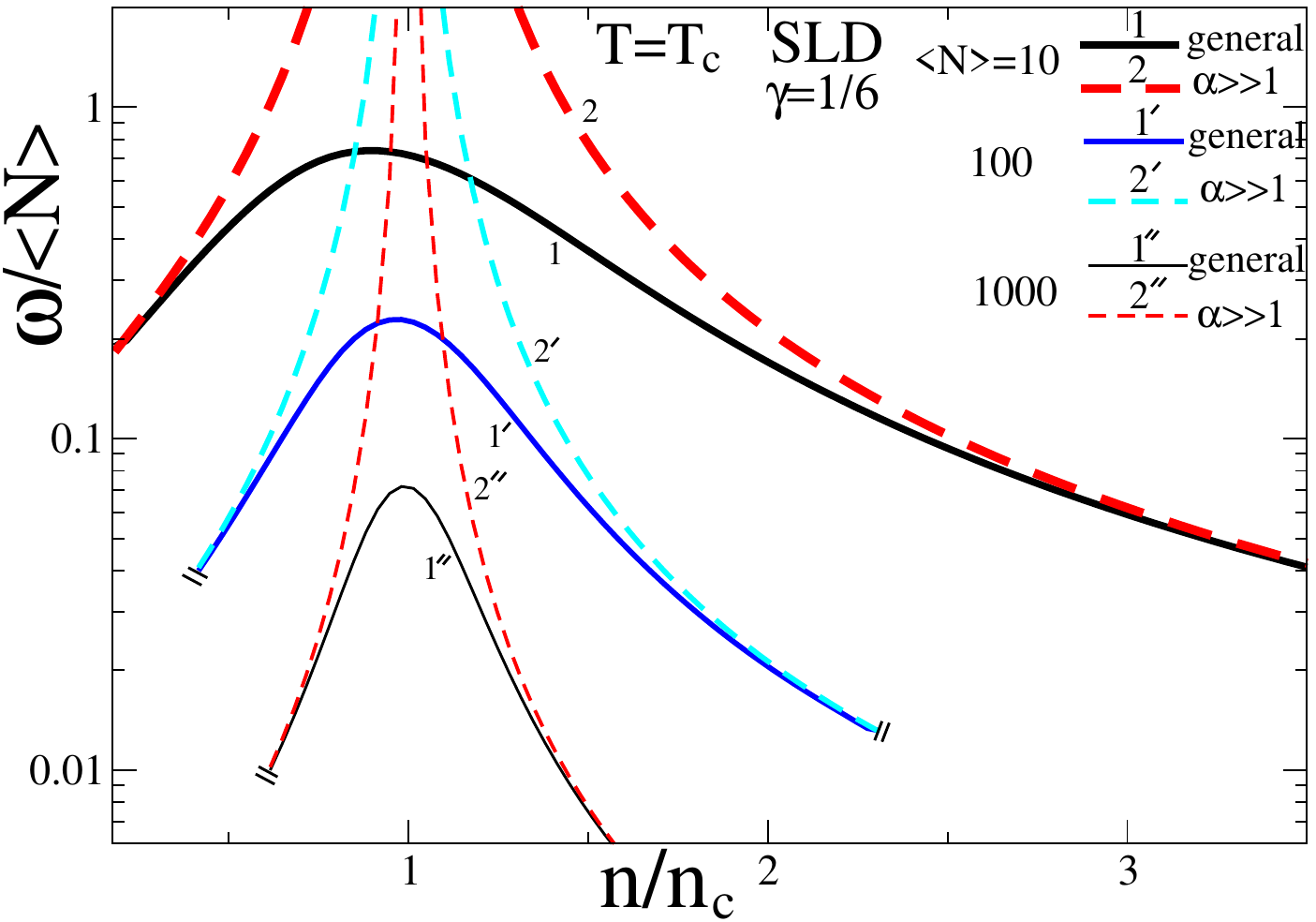}   
\end{center}

  \caption{ The same as in Fig.~\ref{fig5} but for the SLD effective interaction
     with the
     same parameters [Eq.~(\ref{Skyrpar}) and
       critical values of Eq.~(\ref{SkyrCP-1}) and Table \ref{table-2}] as in
     Sec.~\ref{sec-4}
     for $\gamma=1/6$.
  }
\label{fig8}
\end{figure}

Figure~\ref{fig5} shows the particle number fluctuations $\omega$
[Eq.~(\ref{Dn2genK}), 
  solid lines] divided by constant
$\langle N \rangle$, $\omega/\langle N \rangle=
\mbox{D}_N/\langle N \rangle^2$, 
  where $\mbox{D}_N$ is the dispersion 
  for the 
  vdW interparticle interaction parameters, critical temperature ($T=T_c$),
  and several typical particle numbers averages.
Their 
asymptotes,
Eq.~(\ref{lim2}), for large $\alpha$
are shown
at the same 
values of
the particle number average $\langle N \rangle$
(short double lines mean
    an interruption of the lines
to simplify the presentation of the figure). See also
Fig.~\ref{fig6} for
the critical
parameter $\alpha$
as function of the particle number density $n/n_c$ at the critical value
of the temperature $T=T_c$ and the same set of values of $\langle N \rangle$. 
Dashed lines are 
the traditional approach (\ref{lim2}) for $\alpha \gg 1$, valid far from the
critical point (Fig.~\ref{fig5}).  This (traditional)
approach is related to
the second-order power expansion of the free energy $F(\rho)$
over difference $\langle \rho\rangle - n$,
in Eq.~(\ref{Fexp}) for the fixed temperature $T$
at the critical point, $T=T_c$ (see Sec.~\ref{sec-3}).
The dashed lines present the second-order asymptote of the
generalized formula
(\ref{Dn2gen}) at $\alpha \gg 1$.
This traditional
result is the same as that of Eqs.~(\ref{FL-press}) and (\ref{lim2})
for the fluctuations
$\omega$, shown in Figs.~\ref{fig3}(a) and \ref{fig4}(a) as
a curve
for the pure vdW
approach neglecting the quantum effects
but with another normalization.
The normalization of the dispersion
$\mbox{D}_N$ in Fig.~\ref{fig5}
is taken as $\langle N \rangle^2$ for a uniform comparison at different
effective distances $\alpha$ from the CP. There is clearly seen
a divergence of this asymptotic ($\alpha \gg 1$; see dashed lines)
approach
at the
critical point as in
Fig.~\ref{fig3} (a). 
As seen from 
Fig.~\ref{fig5}, one obtains
a maximum of the finite small value 
near the critical 
density value, $n \approx n_c$. This maximum 
    in the dependence
on the particle number density $n$,
$\omega/\langle N \rangle \approx D_N/\langle N \rangle^2$,
near the critical temperature $T \approx T_c$,
  monotonically decreases rapidly with
increasing particle number average $\langle N \rangle$, 
  in contrast
  to an increasing behavior of the dispersion $\mbox{D}_N$
  [Eq.~(\ref{FL-gendef})]. Notice that
our analytical calculations shown in Fig.~\ref{fig5} are in a qualitative
agreement 
with the numerical results presented in Fig.~11 of
Ref.~\cite{AB00}. These results were obtained by using the
numerical statistical percolation model of the phase transitions
\cite{PercolinMod}. We should only take into account that the
second variance in Ref.~\cite{AB00} is related to the dispersion
$\mbox{D}_N$, i.e., the fluctuation $\omega/\langle N \rangle$ in
Fig.~\ref{fig5}, multiplied by $\langle N \rangle^2$. 

    Figure \ref{fig7} presents contour plots for the fluctuations $\omega$
    over the particle number average $\langle N\rangle$,
    i.e., the quantity
    $\omega/\langle N\rangle$. 
    In the upper plot (a)  we show the improved results of calculations,
    according
      to Eq.~(\ref{Dn2genK}),
      while in the bottom plot (b) we consider the
      limit of Eq.~(\ref{Dn2genK}),
        Eq.~(\ref{lim4}),
at $\alpha \ll 1$.
      As seen from these two plots, the results are similar in both 
   panels near the
critical point. We find in both plots a final maximum at a finite
particle number average $\langle N\rangle$, in contrast to
another limit result,  Eq.~(\ref{lim2}),
shown in Fig.~\ref{fig3}(a). It is convenient to normalize the fluctuations
as $D_N/\langle N \rangle^2$ because
the dispersion $D_N$ is  of the order
of
$\langle N\rangle^2$ near the critical point ($\alpha \ll 1$).
This is in contrast to the results
for the fluctuations, valid far from the critical point ($\alpha \gg 1$) where
$D_N$ is of
the order of $\langle N\rangle$, as usual in 
    the standard statistical physics
\cite{LLv5}.

\begin{figure}
  \begin{center}
   \includegraphics[width=8.5cm,clip]{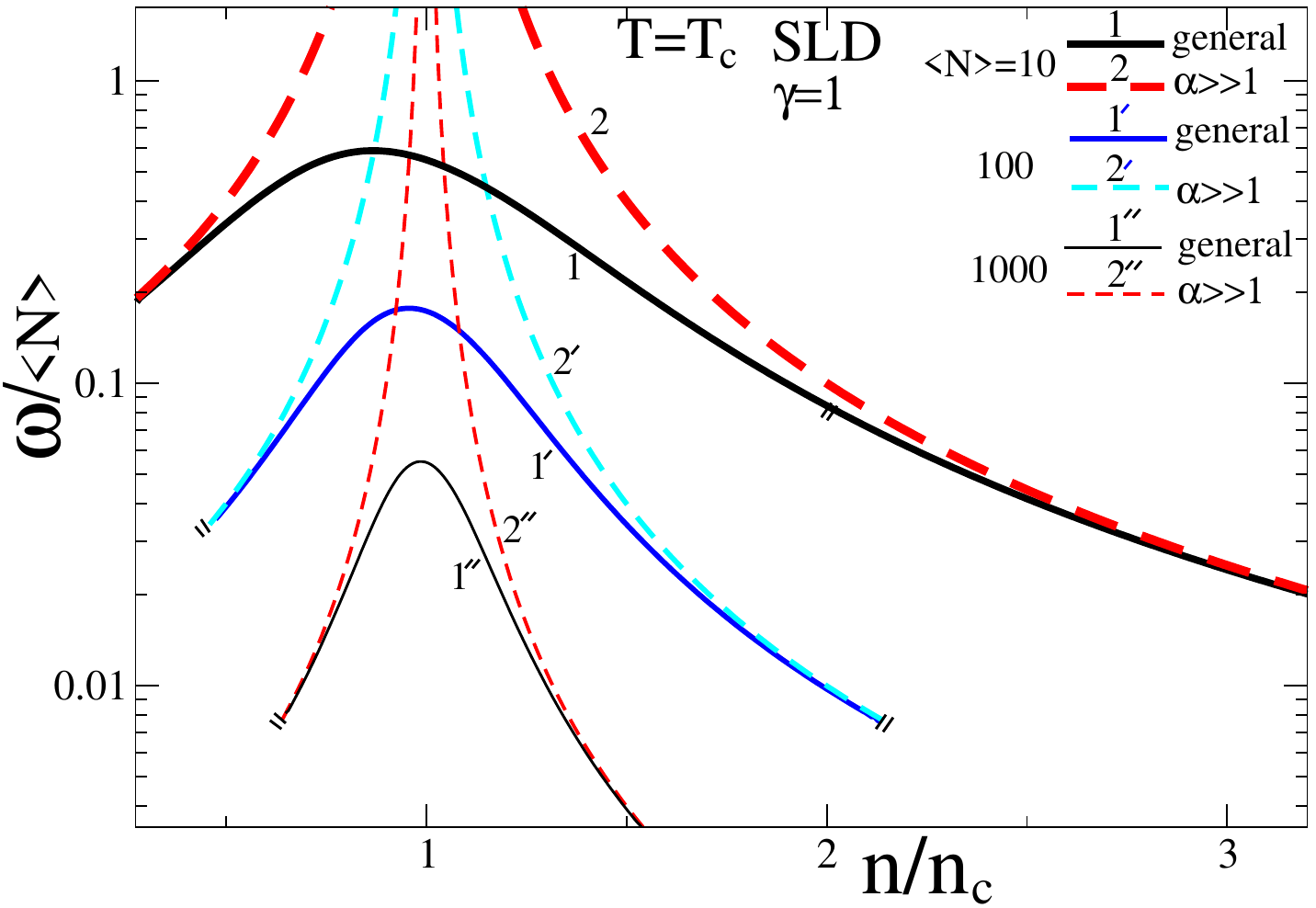}   
\end{center}

  \caption{ The same as in Fig.~\ref{fig8} but for the
    SLD interaction with the
    parameters of Eqs.~(\ref{Skyrpar}) and (\ref{SkyrCP-1}) at $\gamma=1$
    (Table~\ref{table-2}).
    }
\label{fig9}
\end{figure}

\begin{figure}
  \begin{center}
   \includegraphics[width=8.5cm,clip]{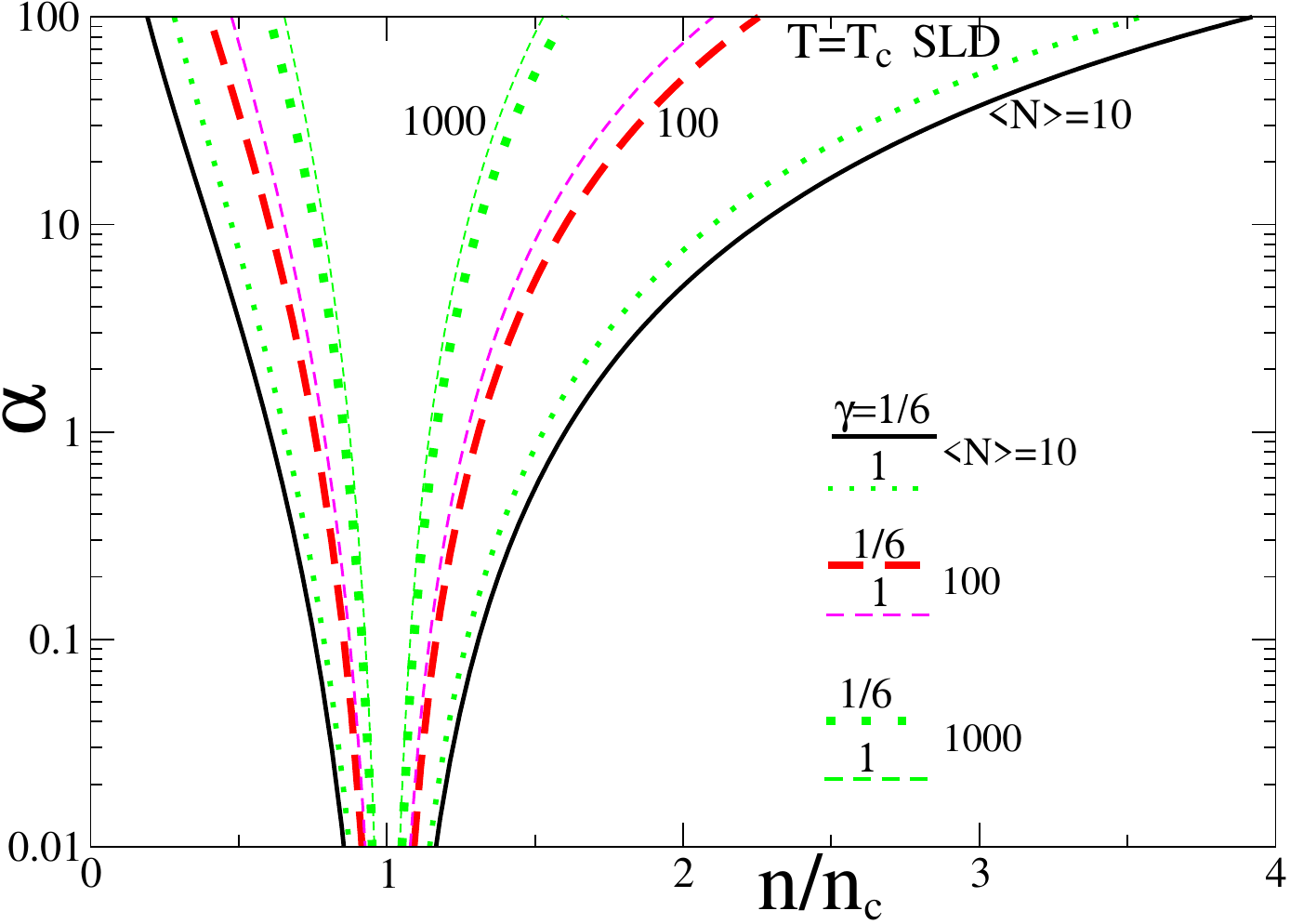}   
\end{center}

  \caption{The same as in Fig.~\ref{fig6} but for the SLD interaction
    with $\gamma=1/6$ and $1$.
  }
\label{fig10}
\end{figure}

Figures~\ref{fig8} and \ref{fig9} show qualitatively the same
   fluctuations, $\omega/\langle N \rangle$,  as in
Fig.~\ref{fig5}, but for the SLD
interaction with parameters $\gamma=1/6$ and $\gamma=1$, respectively;
see also Fig.~\ref{fig10} for the critical parameter $\alpha$
as function of the particle number density $n/n_c$
for the SLD case at both values of $\gamma$.
The difference
between the vdW and SLD cases
is only in a slightly greater
asymmetry of the vdW curves  and their small deflections of the maxima 
from the critical point.
We may note also 
slightly larger
values at maxima in
Figs.~\ref{fig8} and \ref{fig9}, as compared with the
vdW results presented in Fig.~\ref{fig5} for the same particle number
averages $\langle N \rangle$. The
same qualitative agreement with
the results of Ref.~\cite{AB00} was found as for the vdW
interparticle interaction,
mentioned above.

\section{Summary}
\label{sec-sum}

The generalized particle number fluctuations $\omega$ are derived
for an isotopically 
symmetric nuclear matter within the Smoluchowski Einstein
statistical theory. 
This more general result is obtained by using
the fourth-order expansion of the free energy $F(\rho)$
over small difference of the particle
number density $\rho$ from 
its average $n$, and including the second-order terms.
{Thus, we found the fluctuation $\omega$ as a function of the
dimensionless  parameter
$\alpha\propto \mathcal{K}^2 \langle N \rangle/n^2 T
\mathcal{K}^{\prime\prime}$, where $\mathcal{K}$ and
$\mathcal{K}^{\prime\prime}$ are the isothermal
incompressibility and its second derivative at a given temperature $T$.
In} the limit of large $\alpha$, $\alpha \gg 1$, we derived the
traditional asymptotic expression
for the fluctuations $\omega$, $\omega \propto 1/\mathcal{K}$. This result
is
equivalent to that obtained early by 
the second-order power
expansion of the free energy $F(\rho)$ over the particle number
density difference $\rho-n$,
where $n$ is the average of $\rho$.
For small values of $\alpha$ near the critical point, $\alpha \ll 1$,
one finds another known finite asymptotic expression of $\omega$.
This expression is improved locally near this point 
  at finite particle number
averages $\langle N \rangle$. Such an asymptote was derived early
by using the fourth-order power
expansion of the free energy $F(\rho)$ over small
$\rho-n$ but neglecting
the second-order term which is zero at the critical point.
    We found that the values of $\alpha$ determine the effective distances
from the
critical point where one can apply these well-known asymptotes.
These results are obtained for any interparticle interactions.
In addition, these two asymptotes were studied in detail by
using the specific vdW and SLD
interactions as simple examples.

Equations of state obtained
within the quantum van der Waals (QvdW) and
 Skyrme local density (QSLD)
approaches
were used to study analytically
the  
particle-number fluctuation $\omega$, first
by the traditional calculations. These analytical calculations
were performed 
in terms of the isothermal
incompressibility $\mathcal{K}~$, $\omega \propto 1/\mathcal{K}$, 
in the vicinity of the critical point in isotopically symmetric
nuclear matter. The 
expressions for the fluctuations $\omega$ are obtained accounting for the
leading first-order
corrections using the quantum statistics expansion 
over the  small
parameter, $\propto \varepsilon \propto \hbar^3n/g(mT)^{3/2}$, 
in the QvdW model
and that ($\varepsilon $) in the QSLD model. 
A simple and explicit dependence of the particle number fluctuations $\omega$
on the system parameters,
such as the
particle mass $m$, degeneracy factor $g$, and interaction parameters,
$a$ and $b$
for the QvdW and $a^{}_{\rm Sk}$, $b^{}_{\rm Sk}$, and $\gamma$ for the
 QSLD approaches,
is demonstrated at the
 first order of this expansion.
 Such an analytical dependence
 on the particle mass $m$ and degeneracy factor $g$
is absent within the classical vdW and SLD approximations. 
The quantum correction effects, which are quite significant to
obtain the CP parameters of the
nucleon
matter, 
appear to be small for the fluctuations $\omega$.
    They lead to a notable 
    asymmetry of the $\omega(T,n)$ values in the $T-n$ plane as function of
    temperature $T$
    for both discussed models.
     In this respect, the temperature dependence of the
    fluctuations $\omega$ is especially
pronounced for all these approximations.

We 
derived the analytical expressions for 
the fluctuations $\omega$ in terms of the incompressibility $\mathcal{K}$
near the critical point 
 as functions of the
distances from the CP, in units of $ T_{\rm c}$ and $n^{}_{\rm c}$.
 For the temperature
 $T$ behavior of the fluctuations $\omega$
 at constant critical density, $n=n^{}_{\rm c}$,
 one obtains $\omega \propto (T-T_{\rm c})^{-1}$
 with the critical index -1 for the order parameter $T-T_c$ of the
  Landau theory of phase transitions.
 The particle number density $n$ dependence of $\omega$
 at $T=T_{\rm c}$ has another critical index -2,
 $\omega \propto (n-n_{\rm c})^{-2}$,
 for the order parameter $n-n_c$.
The temperature behavior of the fluctuations $\omega$
was obtained to be qualitatively the same for the QvdW and  QSLD approaches but
with  slightly  
different slope coefficients. They are in good agreement with more accurate
numerical calculations for the QvdW case. 
 To our knowledge, there are no numerical
   results 
     for the fluctuation
    slope constant
    in the QSLD case.
     The QvdW  
     density dependence 
    near the CP is
    essentially different from that of the SLD model
    by the slope coefficient. This is
    in contrast to the slope coefficients in
     temperature dependence of the fluctuations $\omega$. 
We found good qualitative and quantitative
agreement between these analytical results and those accounting for a
high-order
derivative expansion near the critical point which 
were suggested
by Tolman and Rowlinson.

In line with 
the accurate
traditional numerical calculations
of the particle number fluctuations $\omega$ in terms of the
incompressibility $\mathcal{K}$,
$\omega \propto 1/\mathcal{K}$, we found analytically
an expected huge bump near the critical point. The obvious reason is
the 
divergence
in the zero incompressibility limit, $\mathcal{K}\rightarrow 0$, at the
CP, for all
compared
approaches to the incompressibility $\mathcal{K}$.
The results are similar to those of the approximate first-order
analytical and 
    more accurate numerical calculations 
  realized 
    with and without using the
    expansion of the incompressibility near the CP at zeroth-
    (vdW or SLD)
    and first-order (QvdW
    or QSLD) approaches
over a small parameter of the quantum statistics, respectively.
Several leading high-order derivative approximations to the
incompressibility $\mathcal{K}$
    were analyzed near the
    critical point. The convergence of the 
    simplest explicitly given analytical results 
    for the
    fluctuations $\omega$  near the critical point
    to their 
    approximations,
    suggested by Tolman and Rowlinson in
    Refs.~\cite{TR38,RJ58}, was found
for the isothermal incompressibility $\mathcal{K}$.
This is expected because, as is well known, the traditional
calculations of particle number
fluctuations $\omega$ in terms of the
incompressibility diverge at the critical point for infinite
nuclear matter. Therefore, these
results cannot be applicable 
 in a close distance 
from the CP. 
They lead to indetermination of the
corresponding averaged particle numbers, 
which are defined up to their fluctuations,
in the equation of state.  The well-known reason
is that the derivation of these
particle number fluctuations $\omega$ in terms of the isothermal
susceptibility, or the
incompressibility $\mathcal{K}$, from the original definition through the
moments of the Gibbs distribution over particle
numbers in the grand canonical ensemble fails  if fluctuations
    are not small.
This is common for any
used interparticle (vdW and SLD) interactions. 
These results of the fluctuation calculations are 
    weakly dependent on the quantum statistics corrections.

       We analyzed the particle
    number  fluctuations $\omega$,
    improved near the critical point, for 
        a finite particle-number piece of  nuclear matter.
   We took into account the additional fourth-order terms in
   expansion of the
   free energy $F(\rho)$ in powers of small difference
   between the density $\rho$ and its average $n$
    beyond the quadratic approximation of the traditional
   classical-fluctuations
   theory, but along with the quadratic terms.
       Using  the vdW and SLD interparticle interaction approaches
    and neglecting small quantum statistical effects,
      we obtained analytically 
    finite values of the particle number
    fluctuations $\omega$
    near the critical point at any finite particle number average
    $\langle N\rangle$. This is  in contrast to the 
   traditional divergent calculations in terms of the
   incompressibility
            or particle number density
            susceptibility. As shown in our calculations, the fluctuations
            $\omega$, divided by the particle number averaging
            $\langle N\rangle$, have a relatively small finite
            maximum near the critical point.
            This maximum of the particle number fluctuations,
            $\omega/\langle N\rangle$, 
            decreases with increasing particle number
            average $\langle N\rangle$, having the zero limit when
            $\langle N\rangle$ goes to the
              infinite. For the dispersion (or variance)
              $\mbox{D}_N$, one respectively finds the increasing
              dependence on
              $\langle N\rangle$, in agreement with the numerical
              results obtained earlier by the percolation model of 
            phase transitions.
                A range of the critical point  vicinity,
                where the traditional ($\alpha \gg 1$) results for
                fluctuations,
                $\omega \propto 1/\mathcal{K}$,
            cannot be applied, decreases with increasing particle
            number average $\langle N \rangle$.  The transition range between
            the two asymptotes, $\alpha \gg 1$ and $\alpha \ll 1$,
            is smaller the larger the  value of
            $\langle N \rangle$ is
            for significantly large values of $\langle N \rangle$.

            As perspectives, 
                we are going to develop the
            Smoluchowski Einshtein
            method to higher-order expansions over the order parameter
            $\rho-n$ than the present fourth-order approach for studying the
            phase transitions.
We will study also
the fluctuations near the critical point in terms of 
moments of the
statistical level density 
    by using another alternative microscopic-macroscopic approach
for
finite Fermi systems \cite{MS21npa,MS21prc,MS21ijmpe,MS22FLT}.
The improved saddle-point method
\cite{Fe62,Fe77,mafm,maf,MY11,MK16,MA17} for analytical calculations of the 
inverse Laplace integrals for the level density near the critical point
will be used to remove the divergences.
Then,
we will calculate
the level-density moments averages
over the particle number and other variables by using the
initial definition for the
corresponding statistical fluctuations.
 Our 
derivations within 
the vdW and SLD forces 
can be straightforwardly
extended to other types of interparticle interactions,  in particular,
to more general and more realistic statistical
nuclear approaches. In particular, our derivations might be
    extended to account for the isotopic proton-neutron asymmetry. 
 We believe that our results of interest also to 
    shed more light on the reasons for 
    the experimental opalescence phenomenon data.

\begin{acknowledgments}
  We gratefully thank  M.I.~Gorenstein and A.S.~Sanzhur
  for many fruitful discussions and suggestions, as well 
  D.V.~Anchishkin, A.~Bonasera, J.~Natowitz, 
  R.V.~Poberezhnyuk, and V.~Vovchenko
for many useful discussions. 
    We thank also very much A.~Haensel
for important help us with
computer facilities used in our calculations. 
A.G.M. thanks very much for the nice hospitality extended
him during staying at the Cyclotron Institute of the Texas A\&M University.
UVG thanks also very much for the nice hospitality during his
staying at the University of Groningen in Netherlands. 
S.N.F., A.G.M., and U.V.G. acknowledge
support in part by the budget program ``Support for the
development of priority   areas of scientific researches,''
a project of the Academy
of Sciences of Ukraine (Code 6541230, No. 0122U000848).  
 A.G.M.  acknowledges also support in part
by the US Department of Energy under Grant No. DE-FG03-93ER-40773. 
\end{acknowledgments}

\appendix

\renewcommand{\theequation}{A.\arabic{equation}}
\renewcommand{\thesubsection}{A\arabic{subsection}}
  \setcounter{equation}{0}

\section{Fluctuations and susceptibility} 
\l{appA}

According to Ref.~\cite{ZM02}, taking the variations
of both sides of
Eq.~(\ref{avpartnumb}) over $\mu$ with the
help of
Eqs.~(\ref{distfun}) and (\ref{partfun}) and changing the order of the
 integrations
over the phase space $\Gamma$ and derivative over the chemical potential $\mu$,
for small first-order variations $\delta \mu$
 in $\mu$, 
 for the particle number 
     fluctuations $\mbox{D}_N/\langle N \rangle$, where
     $\mbox{D}_N=\langle(\Delta N)^2\rangle$ is the particle number dispersion
 normalized to $\langle N \rangle$, one obtains 
\be
  \omega (T,n) =
 \frac{T \chi}{n}~,
\quad \chi=\left(\frac{\delta n}{\delta \mu}\right)^{}_T~, 
  \l{FL-sus}
\ee
where  
$n=n(T,\mu)$ is the particle number density average in the
grand canonical ensemble. Notice that this result is the same as that found
in Ref.~\cite{TR38} 
[Eqs.~(\ref{FL-gendef}) and (\ref{FL-MM})] in the second-order approximation
of the Smoluchowski and Einstein fluctuation theory (Sec.~\ref{sec-3}).
Evaluating this linear response $\chi$ far
    from the critical point as
    $\left(\delta n/\delta \mu\right)^{}_T\sim n/\mu$, one finds small
    fluctuations   $\omega (T,n)\sim T/\mu$   if $T/\mu \ll 1$, 
        i.e., for relatively small temperatures.
    
In Eq.~(\ref{FL-sus}), the 
variation derivative is the isothermal susceptibility $\chi $.
    Assuming, again, small relative fluctuations
   with respect to the average particle number, at 
    the linear (first-order) variations, 
   we can restrict ourselves to
the linear response function (linear susceptibility),
\be\l{chi1}
\chi_{}^{(1)}=(\partial n/\partial \mu)^{}_T~.
\ee
Within this linear approximation, one has 
 explicitly
\begin{equation}
 \omega \approx  \omega^{(1)}\left(T,n\right)
         =\frac{T}{n}\left(\frac{\partial n}{\partial \mu}\right)^{}_T~.
 \label{FL-1sus}
\end{equation}
   The linear response $\chi$ [Eq.~(\ref{FL-sus})-(\ref{FL-1sus})]
 diverges at the critical point, in contrast to its derivations.
 As shown in Appendix \ref{appB} under the same condition
 of small fluctuations, one obtains from
    Eq.~(\ref{FL-sus}) [in particular, from Eq.~(\ref{FL-1sus})]
    the well-known expression (\ref{FL-press}) [or Eq.~(\ref{incompexp1})]
    for the fluctuations, normalized to  $\langle N \rangle$, 
    in terms of the
    isothermal incompressibility $\mathcal{K}$
    \cite{TR38,RJ58,LLv5,TK66,IA71,BR75,AC90,ZM02}.

 Let us consider variations of the relationship (\ref{avpartnumb}) over the
 chemical potential $\mu$, taking into account high-order variations,
 for instance
 second-order ones. 
 We will still take these variations
 at constant
  temperature $T$, i.e., consider nonlinear (second-order)
 isothermal susceptibility $\chi^{(2)}$. 
  Equation~(\ref{FL-sus}) is 
  valid for any order of the
 variation derivative (nonlinear susceptibility), 
 but now
  one can specify it 
 for the second-order fluctuations $\omega^{(2)}$.
 Taking immediately the  variations over $\mu$ up to the 
     second order
 at $T=const$ in Eq.~(\ref{avpartnumb}), one obtains
 high 
 (second) order corrections to Eq.~(\ref{FL-1sus}). 
   Equation~(\ref{FL-1sus}) is named usually
 the second cumulant of the averaged 
 Gibbs distribution function, Eq.~(\ref{distfun}), averaged over
   the phase space.
  The dispersion $\delta^{(2)}(\langle N \rangle)$, 
 taking into account up to the 
 third
 cumulant moment of the averaged Gibbs distribution, take the form
 \be\l{2ndordervar}
 \frac{T}{\langle N \rangle}\delta^{(2)}(\langle N \rangle)=
 \omega^{(1)}~
 (\delta \mu)^{1} 
 +\frac{1}{2T} \omega^{(2)} 
 (\delta \mu)^2 +\ldots~,
 \ee
 where 
  $\omega^{(2)}$ is the so-called kurtosis.
 It can be normalized by $\langle N^2\rangle$, in analogy with 
 $\omega^{(1)}$, Eq.\ (\ref{FL-1sus}) (see Ref.~\cite{marik}): $\omega^{(2)}=
  \left(\langle N^3\rangle -\langle N\rangle^3\right) /\langle N^2 \rangle$~.
 Similarly, one can obtain the third-order moment (or 
 third cumulant) 
 of the averaged
 Gibbs distribution, Eq.~(\ref{distfun}). This third-order moment 
is coming from the third-order
variations of the average $\langle N\rangle$, Eq.~(\ref{avpartnumb}),
over the chemical potential
 $\mu$, 
 and so on.
 This allows us to go beyond the restrictions of the 
 first-order cumulant 
 fluctuations $\omega^{(1)}$, shown explicitly in Eq.\ (\ref{FL-1sus}).
  Namely, this is beyond the first variation derivative for the
 susceptibility $\chi$: ~ linear susceptibility $\chi^{(1)}$, Eq.~(\ref{chi1}).
  The expression (\ref{FL-sus})
 for the fluctuation $\omega$ of the particle number is more general. 
 However, it is still singular exactly at the CP where the linear susceptibility
 $\chi^{(1)}$ (\ref{chi1}) is infinity in the sum (\ref{2ndordervar}).

\renewcommand{\theequation}{B.\arabic{equation}}
\renewcommand{\thesubsection}{B\arabic{subsection}}
  \setcounter{equation}{0}

\section{Derivations of the classical particle-number fluctuations}
\l{appB}

Within the canonical ensemble, 
    one can use the free energy 
    $F(T,N,V)$, Eq.~(\ref{F}), as a
    characteristic thermodynamic function of the 
    temperature $T$, particle number 
    $N$, and volume $V$. Assuming the thermodynamic limit
    condition for our infinite
system, one can express $F$ in terms of that per particle \cite{LLv5},
\be\l{f}
F(T,N,V)=Nf(T,\tilde{v})~,
\ee
 where $\tilde{v}$ is the volume per particle,
\be\l{vtilde}
\tilde{v}=\frac{1}{n},~~~ n=N/V~.
\ee
For the pressure $P$ and chemical potential $\mu$, one has
\be\l{P}
P=-\left(\frac{\partial F}{\partial V}\right)^{}_T=
-\left(\frac{\partial f}{\partial \tilde{v}}\right)^{}_T~,
\ee
and
\be\l{mu1}
\mu=\left(\frac{\partial F}{\partial N}\right)^{}_T=f - \frac{1}{n}
\left(\frac{\partial f}{\partial \tilde{v}}\right)^{}_T~.
\ee

Taking the first variation of Eq.~(\ref{mu1}) over particle number
density $n$ through
the relationship (\ref{vtilde}), one obtains
\be\l{dmudn}
\delta \mu=
\frac{1}{n^3}\left(\frac{\partial^2 f}{\partial \tilde{v}^2}\right)^{}_T~
\delta n~~.
\ee
Therefore, one finds
\be\l{dndmu}
\left(\frac{\partial n}{\partial \mu}\right)^{}_T=
\frac{n^3}{\left(\partial^2 f/\partial \tilde{v}^2\right)^{}_T}~.
\ee
According to Eq.~(\ref{FL-1sus}) and Eqs.~(\ref{dndmu}), (\ref{P}), and
(\ref{vtilde}),
one arrives at Eq.~(\ref{FL-press}).

Note that the same result can be obtained more easily 
by using the
Jacobian (linear) transformations 
\cite{LLv5}
\be\l{Jactrans1}
\left(\frac{\partial n}{\partial \mu}\right)^{}_T=\frac{D(n,T)}{D(\mu,T)}=
\frac{1}{D(\mu,T)/D(n,T)}~
\ee
and
\be\l{nGCE}
n=\left(\frac{\partial P}{\partial \mu}\right)^{}_T =~
\frac{D(P,T)}{D(\mu,T)}~.
\ee
Therefore, substituting 
Eqs.~
(\ref{Jactrans1}) and (\ref{nGCE})  into Eq.~(\ref{FL-1sus})
for the particle number fluctuations $\omega$,
one can carry out cancellation in ratios of the denominator 
by using the Jacobian properties. Finally, 
 once again one obtains Eq.~(\ref{FL-press}).

Note that these derivations, based on the first derivative
transformations, fail near the
critical point because of the divergence of fluctuations due to zeros
in the denominators. 
Therefore, strictly speaking, Eq.~(\ref{FL-press}) cannot be used
in the close vicinity of the critical point [see Eq.~(\ref{CP-eq})],
in contrast to
the fluctuation formula; 
see, e.g., Eq.~(\ref{Dn2genK})
  obtained in Sec. \ref{sec-5} from the 
moments of the averaged
Gibbs distribution.

\renewcommand{\theequation}{C.\arabic{equation}}
\renewcommand{\thesubsection}{C\arabic{subsection}}
  \setcounter{equation}{0}
\section{
   Analytical critical-point results within the QvdW and QSLD models}
\label{appC}

\subsection{
  The van der Waals  model with quantum-statistics corrections}
\label{appC1}

 Following Refs.~\cite{FMG19,FMG22}, we introduce
 a small quantum statistics
parameter $\delta$ of
expansion of the pressure $P(T,n)$, accounting for the vdW interaction
in terms of the vdW
attractive parameter $a$, and repulsive exclusion-volume parameter $b$.
 For the Fermi statistics, one has Eq.~(\ref{del}) for $\delta$.
  Up to the  first leading quantum statistics
      corrections over $\delta$
    to the
 vdW model, one has
\be\label{Pvdw-n}
 P_{\rm W}(T,n) = \frac{nT}{1-bn}\left[1+\delta+
    \mbox{O}\left(\delta ^2\right) \right]
   -a\,n^2.
   \ee
   It was shown in Refs.~ 
   \cite{FMG19,FMG22} that
   at small $\delta$
  the expansion of the pressure $P_{\rm W}(T,n)$
  over powers of $\delta$ becomes rapidly convergent
  to the accurate results
      for sufficiently large temperature $T$ and small
   particle-number density $n$. Therefore,
   even the first-order
    terms provide
    already a good approximation.
 The first  quantum-statistics corrections in
     Eq.~(\ref{Pvdw-n})
      increase
      with the particle number density $n$ and decrease 
       with the increase of the system temperature $T$,
      particle mass $m$, and degeneracy factor $g$.
      A new feature of
quantum statistics
effects in the system of  particles with the
    vdW
interaction is the additional
factor $(1-bn)^{-1}$ in the
correction $\delta$ [Eq.~(\ref{del})] 
with respect to the
ideal gas case. 
 Thus, the
quantum statistics effects
  become stronger due to 
  the repulsive
interaction between particles.

The first-order equation of state [Eq.~(\ref{Pvdw-n})]
within the quantum vdW (QvdW) model
describes the corresponding 
liquid-gas
phase transition. The critical point (CP) of this transition satisfies
the 
equations of (\ref{CP-eq}) \cite{LLv5}.
Using Eq.~(\ref{Pvdw-n}) in the first
approximation
over $\delta$,
one derives
from Eq.~(\ref{CP-eq})
the
system of two equations
for the CP
parameters $n_c$ and $T_c$
at the same 
 corresponding order.
 The solutions of this
 system
 in the same first-order  
 approximation
over $\delta$ have
 the form
\bea
 & T_c^{(1)} ~ \cong
  T_c^{(0)}\left(1 - 2 \delta^{}_0  \right)
 ~,\nonumber \\
& n_c^{(1)} \cong  n^{(0)}_c\left(1 - 2\delta^{}_0  \right)
~.  \label{CP-1}
  \eea
In Eq.~(\ref{CP-1}), 
the values $ T_c^{(0)}$ and $ n_c^{(0)}$ are the CP parameters of the classical
vdW model
with the pressure [Eq.~(\ref{Pvdw-n}) at  $\delta=0$]
\be\label{vdW}
 P^{(0)}_{\rm W}(T,n) = \frac{nT}{1-bn}
   -a\,n^2~.
   \ee
   These CP values are the zero-order approximation
   in the QvdW, $\delta=0$:
\bea
  & T_c^{(0)}=\frac{8a}{27b}\cong 29.2~{\rm MeV}~, ~~~
  n_c^{(0)}=\frac{1}{3b}\cong 0.100~{\rm fm}^{-3}~,\nonumber \\
& P_c^{(0)}=\frac{a}{27b^2}\cong 1.09~{\rm MeV}\cdot {\rm fm}^{-3}~. \label{CP-0}
  \eea
The constants $a$ and $b$ of the  QvdW model,
$a>0$ and $b>0$, are responsible for 
attractive and repulsive
interactions between particles, respectively.
We will 
compare our analytical
first-order results for fluctuations with those of more accurate numerical
 calculations \cite{roma,satarov1,roma2,oleh20,St21-1,St21-2,Kuzn21}.
Therefore, 
as in Refs.~\cite{marik,vova,satarov,FMG19,FMG22},
we fix the model parameters $a$ and $b$ using the ground state
properties  of  isotopic 
symmetric nuclear matter
(see, e.g., Ref.~\cite{bethe}): at $T=0$ and
  $n=n_{0}= 0.16~\mbox{fm}^{-3}$, one requires
$P=0$ and the binding energy per nucleon $\varepsilon(T=0,n=n_0)/n_0=- 16$~MeV.
From the above requirements,
one finds
\be\l{ab}
a=329.8\, \mbox{MeV} \cdot \mbox{fm}^3 , \;\;\; b=3.35\, \mbox{fm}^3~.
\ee
The parameter $\delta^{}_0$
in Eq.~(\ref{CP-1}) 
is given by Eq.~(\ref{del}),
 taken at the CP
 of the zero-order approximation (\ref{CP-0}),
    i.e., at $n=n_c^{(0)}$ and $T=T_c^{(0)}$,
    $\delta^{}_0=\delta\left(T=T^{(0)}_c,n=n_c^{(0)}\right)$.
Substituting Eq.~(\ref{CP-1}) 
for the results of the
 corresponding critical temperature, $T_c^{(1)}$, and density,
$n_c^{(1)}$, 
into the equation of state
[Eq.~(\ref{Pvdw-n})], at a given
perturbation order, one can calculate the
CP pressure $P_{c}^{(1)}$
at the same order, $P_c^{(1)}=P_{\rm W}(T=T^{(1)}_c,n=n_c^{(1)})$
    [Eq.~(\ref{Pvdw-n})].  
Notice that the temperature $ T_c^{(1)}$ and
density $ n_c^{(1)}$ are decreased for Fermi statistics with respect to
$ T_c^{(0)}$ and  $ n_c^{(0)}$, in contrast to
the opposite behavior 
for Bose particles.

 \subsection{
   The
   Skyrme local-density model
   with quantum statistics corrections}
 \label{appC2}

The pressure function of the quantum-statistics
Skyrme local-density 
(QSLD) model
\cite{satarov}, after some transformations,
 can be presented
as \cite{FMG22}
\be\l{PQSkyr}
P_{\rm Sk}(T,n)=
nT\left(1 + \varepsilon\right) 
- a^{}_{\rm Sk}n^2 + b^{}_{\rm Sk}n^{\gamma +2}~,
\ee
where
$a^{}_{\rm Sk}$, $b^{}_{\rm Sk}$, and $\gamma$ are
 interaction constants of the
 QSLD parametrization \cite{satarov}.

  Within the QSLD
  approach, one can
 consider the critical points for a
first-order liquid-gas phase transition, for instance,
for pure nucleon 
matter. 
The critical point (CP) for
     the QSLD model obeys
the
same equation (\ref{CP-eq}) but
 with the quantum-statistics Skyrme local-density
 pressure,
$P=P_{\rm Sk}(T,n)$ [Eq.~(\ref{PQSkyr})]. 
Solving the system of equations  
(\ref{CP-eq}) 
with the equation of state (\ref{PQSkyr}) in the first-order approximation
over $\varepsilon$, Eq.~(\ref{del}),
one obtains \cite{FMG22}
\bea
& T_{{\rm Sk},c}^{(1)} ~ \cong
  T_{{\rm Sk},c}^{(0)}\left(1 - 2 \varepsilon^{}_{0}\right)
~,\nonumber \\
  & n_{{\rm Sk},c}^{(1)} \cong  n^{(0)}_{{\rm Sk},c}
  \left(1-\frac{2 \varepsilon^{}_0}{\gamma+1}\right)~.
  \label{SkyrCP-1}
  \eea
In Eq.~(\ref{SkyrCP-1}), the
temperature  $T_{{\rm Sk},c}^{(0)}$ and density $n_{{\rm Sk},c}^{(0)}$ are the
solutions
of equations
 [see Eq.~(\ref{CP-eq}) with the QSLD pressure (\ref{PQSkyr})]
at zero-order perturbation, $\varepsilon=0$: 
\bea\l{SkyrCP-0}
& T_{{\rm Sk},c}^{(0)}=\frac{2\gamma a^{}_{\rm Sk}n_{{\rm Sk},c}^{(0)}}{\gamma+1}~,
\nonumber\\
& n_{{\rm Sk},c}^{(0)}=\left[
  \frac{2a^{}_{\rm Sk}}{b^{}_{\rm Sk}(\gamma+1)(\gamma+2)}\right]^{1/\gamma}~;
\eea
see also Ref.~\cite{satarov0} where another Skyrme parametrization
for the critical temperature and particle number density at zero
quantum statistics
    corrections was used.
    For the parameters $a^{}_{\rm Sk}$ and $b^{}_{\rm Sk}$
    of Skyrme parametrization,
 the degeneracy 
 for nucleon system, $ g=4$, and $~m=938~\mbox{MeV}$,
 one has \cite{satarov} 
 \begin{eqnarray}\label{Skyrpar}
   &  a^{}_{\rm Sk}=1.167~\mbox{GeV}\cdot{\mbox{fm}}^{3},\nonumber\\
   & b^{}_{\rm Sk}=1.475~\mbox{GeV}\cdot{\mbox{fm}}^{3+3\gamma},~~~
   \gamma=1/6~,\nonumber\\
   & a^{}_{\rm Sk}=0.399~\mbox{GeV}\cdot{\mbox{fm}}^{3}, \nonumber\\
   & b^{}_{\rm Sk}=2.049~\mbox{GeV}\cdot{\mbox{fm}}^{3+3\gamma},~~\gamma=1~.
   \end{eqnarray}
  The QSLD parameters
 are chosen by fitting
 the properties of
one component (in our case, nucleons)
at the
    temperature $T=0$.
 
 The value $\varepsilon^{}_{0}$ in Eq.~(\ref{SkyrCP-1})
    is defined
    by Eq.~(\ref{del}) for $\varepsilon$ at $T=T_{{\rm Sk},c}^{(0)}$ and
    $n=n_{{\rm Sk},c}^{(0)}$
    [Eq.~(\ref{SkyrCP-0})].
    For the CP pressure at $\varepsilon=0$, one finds from Eqs.~(\ref{PQSkyr})
 and (\ref{SkyrCP-0}), 
\be\label{SkyrCP-0P}
P_{{\rm Sk},c}^{(0)}=n_{{\rm Sk},c}^{(0)} T_{{\rm Sk},c}^{(0)}
-a^{}_{\rm Sk}\left[n_{{\rm Sk},c}^{(0)}\right]^2
+b^{}_{\rm Sk}\left[n_{{\rm Sk},c}^{(0)}\right]^{\gamma+2}. 
\ee
The first-order pressure, $P_{{\rm Sk},c}^{(1)}$, 
can be 
 straightforwardly
 calculated  from Eq.~(\ref{PQSkyr}) 
  using the
expressions for $T_{{\rm Sk},c}^{(1)}$ and
$n_{{\rm Sk},c}^{(1)}$ [Eq.~(\ref{SkyrCP-1})],
$P_{{\rm Sk},c}^{(1)}=P_{{\rm Sk}}(T=T_{{\rm Sk},c}^{(1)},n=n_{{\rm Sk},c}^{(1)})$
[Eq.~(\ref{PQSkyr})].

\renewcommand{\theequation}{D.\arabic{equation}}
\renewcommand{\thesubsection}{D\arabic{subsection}}
  \setcounter{equation}{0}

\section{More accurate improved fluctuations} 
\l{appD}

Taking into account the quadratic term in the expansion (\ref{Fexp})
along with the second-order term, 
the quantity $(\rho-n)^2$,
which we are
going to average, should be statistically consistent
with the expansion  (\ref{Fexp})
up to fourth order terms in the mean field approximation \cite{TR38,huang}.
Using the denotation $x=\rho-n$
for shortness,
       for $x^2$ one has 
    [see
    Eq.~(\ref{DF})]
    \be\l{sceq}
 x^4 + \mathcal{A} x^2-\mathcal{B}\{\rho\}=0~,
    \ee
    where
    \be\l{ABdef}
    \mathcal{A}=12 F_2/F_4,\quad \mathcal{B}\{\rho\}=24\Delta\{\rho\}/F_4.
    \ee
              Here, $F_2$ and $F_4$ are the derivatives of the free 
         energy $F$ over the
    density $\rho$:
\be\l{Fm}
F_m=(\partial^m F/\partial \rho^m)_{\rho=n}~,\quad m=2,4~,
\ee
    and $\Delta\{\rho\}$ is given by Eq.~(\ref{DF}).
       Equation (\ref{sceq}) is a complicated self-consistent transcendent
    identity
    for $x$ because the
    last term $\mathcal{B}\{\rho\}$ depends on $x=\rho-n$ in a cumbersome way
    through Eqs.~(\ref{ABdef}) and
    (\ref{DF}).
       Taking the statistical
    average over the Gibbs distribution $W_{\rm eq}^{(N)}$, Eq.~(\ref{distfun}),
    in
    the Eq.~(\ref{sceq}) term by term, one has
    \be\l{avsceq}
    \langle x^4\rangle + \mathcal{A} \langle x^2\rangle-
    \langle\mathcal{B}\{\rho\}\rangle=0~,
    \ee
    where
    \be\l{avB}
 \langle\mathcal{B}\{\rho\}\rangle=24 \langle \Delta\{\rho\}\rangle/F_4~.
    \ee
    The angle brackets have the same meaning as in Sec.~\ref{sec-2},
    including averaging over the phase space ${\bf p}$ and ${\bf q}$, and
    over the particle numbers $N$.
      Expanding $\langle x^4\rangle$ over the statistical correlations,
    one can present
    $\langle x^4\rangle$ in terms of the square  $\langle x^2\rangle^2$
    and a density-density correlation term:
    \be\l{correxp}
\langle x^4\rangle=(\langle x^2\rangle)^2 + \mbox{corr. term}~,
    \ee
    where $\mbox{corr. term}=\langle x^4\rangle-\langle x^2\rangle^2$
    is the density-density correlation term.
    In the mean field approximation,
    we may neglect this small density-density
    correlation term in
    Eq.~(\ref{correxp}) because it is due to the residue interaction above
    a mean field.

    It seems that we do not need to take care of the identities
    (\ref{sceq}) and
    (\ref{avsceq})
    in the case when we might be able to solve analytically exactly our 
    problem with the Gibbs averaging, accounting for statistical
    correlations in all orders above the mean
    field approximation. However, simplifying our statistical problem by using
    this mean field approach, one should take care of executing still these
    identities approximately
    with the statistical accuracy of the mean field approximation, i.e., after
    neglecting correlation terms of Eq.~(\ref{correxp}).
   Thus,  at the zero-order
    approximation over these correlations, from Eq.~(\ref{avsceq}) one finds
    the approximately
    closed equation of the consistency condition (\ref{sceq}),
    taken in average,
    for
    $\langle x^2\rangle$ with an accuracy up to such
    correlations:
    \be\l{conseqx2}
\langle x^2\rangle^2 + \mathcal{A} \langle x^2\rangle-
    \langle\mathcal{B}\{\rho\}\rangle=0~.
    \ee
       This equation optimizes our statistical errors when we
         use the mean
           field approximation $W_4$, Eq.~(\ref{W4}), 
         to the Gibbs distribution
         $W_{\rm eq}^{(N)}$, Eq.~(\ref{distfun}), in evaluations of
         $\langle\Delta\{\rho\}\rangle$; see Eq.~(\ref{DF})
         and Refs.~\cite{TR38,huang}.
    Solving Eq.~(\ref{conseqx2}) with respect to
    $\langle x^2\rangle$,
    for a real positive solution, one
    obtains
    \bea\l{x2}
   & \langle x^2\rangle\approx \frac{\mathcal{A}}{2}\left(\sqrt{1 +
      \frac{4 \langle \mathcal{B}\rangle}{\mathcal{A}^2}}-1\right)\nonumber\\
    &=
    \frac{\mathcal{A}}{2}\left(\sqrt{1 +
      \frac{4 \langle\tilde{\Delta}\{\rho\}\rangle}{\alpha}}-1\right)~,
    \eea
    where $\tilde{\Delta}\{\rho\}= \Delta\{\rho\}/T$ is a dimensionless
    quantity.

    It was convenient and constructive in Eq.~(\ref{x2})
    to re-write the variance
    $\langle x^2\rangle$
    by introducing explicitly
    the critical dimensionless parameter
    $\alpha \propto F_2^2/F_4T$:
    \be\l{alp}
    \alpha=\frac{6 (F_2)^2}{T F_4}~.
      \ee
         Another dimensionless parameter is $c^{}_4 \propto F_4/T$.
         These
      two parameters $\alpha$ and $c^{}_4$ were introduced instead
         of the original parameters $F_2$ and $F_4$ of the potential
         difference
    $\Delta\{\rho\}$ [Eq.~(\ref{DF})]. 
    Then, the constant $\mathcal{A}$ in Eq.~(\ref{x2}) can be
    expressed in terms of $\alpha$, Eq.~(\ref{alp}), and
    $c_4$ as
    \be\l{A}
    \mathcal{A}=\sqrt{\frac{\alpha}{c_4}}~,\quad c_4=\frac{F_4}{24T}~.
    \ee
       It is helpful also to use the obvious relationship
           [see Eqs.~(\ref{alp})
    and (\ref{A})]
    \be\l{BA2}
    \frac{\mathcal{A}^2}{\langle \mathcal{B(\rho)}\rangle}=
    \frac{\alpha}{ \langle\tilde{\Delta}\{\rho\}\rangle}~.
    \ee
       Obviously, at the critical point one has $\alpha=0$ because
    $F_2 \propto \mathcal{K}=0 $ if $F_4$ is assumed to be relatively
    finite,
    $F_4 \geq const > 0$. For small parameter $\alpha$, one has 
        effectively
    a small
    distance from the critical point while for large $\alpha$ one finds
    a large distance from the CP in the averaged particle number density
    $n$ for a given temperature $T$. 
    Thus, $\alpha$ is a dimensionless effective measure
    of the distance from the CP in the density-temperature plane.

    So far in this appendix, the angle brackets were defined as the
    statistical averaging
    with the general Gibbs
    distribution $W_{\rm eq}^{(N)}$ of the grand
    canonical ensemble; see Eq.~(\ref{distfun}) for  $W_{\rm eq}^{(N)}$.
    In order 
    to evaluate now approximately the average of the
    dimensionless potential variation
    $\langle\tilde{\Delta}\{\rho\}\rangle$ which appears in
    Eq.~(\ref{x2}),
    we will use the average statistical distribution function
    $W_4$ as a good approximation to the averaged $W_{\rm eq}^{(N)}$, 
    within the mean field approach.
    Then, for the statistical average of $\Delta\{\rho\}$ [Eq.~(\ref{DF})],
    $\langle \Delta\{\rho\} \rangle$,
    one approximately
    has (see Ref.~\cite{TR38} and Secs.~\ref{sec-3} and \ref{sec-5})
    \be\l{DeF}
    \langle \Delta\{\rho\}\rangle=\langle \Delta_2\{\rho\}\rangle +
    \langle \Delta_4\{\rho\} \rangle~,
    \ee
    where
    \bea\l{DeFm}
    & \langle \Delta_2\{\rho\} \rangle=
    \frac{F_2}{2}\int_0^\infty (\rho-n)^2 W_4\mbox{d}\rho~,\nonumber\\
   & \langle \Delta_4\{\rho\} \rangle=\frac{F_4}{24}
    \int_0^\infty (\rho-n)^4 W_4\mbox{d}\rho~,
    \eea
In Eq.~(\ref{DeFm}), $W_4$ is the probability distribution
given by 
Eq.~(\ref{W4}) 
with the normalization condition
(\ref{normcond}).
With Eq.~(\ref{DF}),
   from Eq.~(\ref{DeF}), one writes
    \be\l{DeF2}
    \langle \Delta_2\{\rho\} \rangle=\frac{F_2}{2}\frac{
      \int_0^{\infty} x^2 \mbox{d}x~\exp\left[
        -\frac{1}{2T}\left(F_2 x^2 +
        F_4 x^4/12\right)\right]}
    {\int_0^{\infty} \mbox{d}x~\exp\left[
        -\frac{1}{2T}\left(F_2 x^2 +
        F_4 x^4/12\right)\right]}~,
    \ee
    and
    \be\l{DeF4}
    \langle \Delta_4\{\rho\} \rangle=\frac{F_4}{24}\frac{
      \int_0^{\infty} x^4 \mbox{d}x~\exp\left[
        -\frac{1}{2T}\left(F_2 x^2 +
        F_4 x^4/12\right)\right]}
    {\int_0^{\infty} \mbox{d}x~\exp\left[
        -\frac{1}{2T}\left(F_2 x^2 +
        F_4 x^4/12\right)\right]}~,
    \ee
    where $x=\rho-n$, as above.

    Using Eqs.~(\ref{DeF}), (\ref{DeF2}), and (\ref{DeF4}) for calculations
    of the average of the dimensionless
    potential difference $\tilde{\Delta}\{\rho\}$, one finds more explicit
    expressions in terms of the modified Bessel functions:
    \be\l{tDelF}
    \langle\tilde{\Delta}\{\rho\}\rangle=
    \langle \tilde{\Delta}_2\{\rho\}\rangle
    +\langle \tilde{\Delta}_4\{\rho\}\rangle,
 \ee
where
\bea\l{tDelF2}
&\langle \tilde{\Delta}_2\{\rho\}\rangle \equiv
\frac{\langle \Delta_2\{\rho\}\rangle}{T}\nonumber\\
&=
\frac{\pi}{4\sqrt{2}}\left\{
\left(\alpha+4\right) I_{1/4}\left(\frac{\alpha}{8}\right)-
  \alpha I_{-1/4}\left(\frac{\alpha}{8}\right)
 \right.\nonumber\\
& -\left.\alpha\left[I_{3/4}\left(\frac{\alpha}{8}\right)-
 I_{5/4}\left(\frac{\alpha}{8}\right)\right]\right\}/
K_{1/4}\left(\frac{\alpha}{8}\right)~.
\eea
and
\bea\l{tDelF4}
&\langle \tilde{\Delta}_4\{\rho\}\rangle \equiv
\frac{\langle \Delta_4\{\rho\}\rangle}{T}=
\frac{1}{8}\left[\left(\alpha+2\right) K_{1/4}\left(\frac{\alpha}{8}\right)
  \right.\nonumber\\
&  -\left.\alpha K_{3/4}\left(\frac{\alpha}{8}\right)
  \right]/ K_{1/4}\left(\frac{\alpha}{8}\right)~.
\eea
Here,
$I_\nu\left(z\right)$ and $K_\nu\left(z\right)$ are the
modified Bessel functions of the order $\nu$ [$K_\nu\left(z\right)$
is named also the MacDonald Bessel function]. 
For $\alpha \gg 1$, far from the critical point,
one obtains $\langle\tilde{\Delta}\{\rho\}\rangle \approx
\langle\Delta_2\rangle/T\approx 1/2$; see Eq.~(\ref{difF}). In the case
$\alpha \ll 1$, near the CP, one obtains
$\langle\tilde{\Delta}\{\rho\}\rangle \approx
\langle\Delta_4\rangle/T\approx 1/4$; see also Eq.~(\ref{difF44s})
in the next appendix. 
    
Dividing by $n^2$ the
 left and final right sides of Eq.~(\ref{x2}),
    one arrives at the dimensionless particle-density fluctuations,
    Eq.~(\ref{FL-MM}); see also Eq.~(\ref{Dn2gen}).
  Differentiating the relationship (\ref{p}) between the
pressure $P(\rho)$ and free energy
$F(\rho)$ over $\rho$, and using the conditions of the statistical equilibrium,
one finds the relationships
\be\l{DFDK}
F_2=\frac{\langle N \rangle \mathcal{K}}{n^2}, \quad
F_4=\frac{\langle N\rangle \mathcal{K}^{\prime\prime}}{n^2}~.
\ee
They are useful in the derivations of Sec.~\ref{sec-5};
see Eq.~(\ref{alpABDF}), and asymptotes (\ref{D2lim})
for $\alpha \gg 1$ and 
(\ref{D4limCP}) for $\alpha \ll 1$, neglecting small corrections
of high order in powers
of $1/\langle N\rangle$. In principle, we may take into account the
density-density correlations
by using the standard iteration procedure. However, to calculate the
correlation term in Eq.~(\ref{correxp}) at any given order we have to
specify the interparticle
interaction.

\renewcommand{\theequation}{E.\arabic{equation}}
\renewcommand{\thesubsection}{E\arabic{subsection}}
  \setcounter{equation}{0}

\section{Asympotical fourth-order improved fluctuations} 
\l{appE}

It is useful to present briefly the derivation of the limit $\alpha \ll 1$
neglecting the second-order term of the free energy expansion at
the very
beginning \cite{TR38,RJ58}. In
this case, for
the free energy expansion, one has
from Eq.~(\ref{DF})
    \bea\l{DF44s}
    &\Delta^{(4)}_4\{\rho\}\equiv F(\rho)-F(n)\nonumber\\
    &=
       \frac{1}{24}
    \left(\frac{\partial^4 F}{\partial \rho^4}\right)^{}_{\rho=n}
    \left(\rho-n\right)^4. 
    \eea

    As shown in Appendix \ref{appD}, in the mean field approximation,
      i.e.,
    at the zero-order density-density correlations, 
     one finds
             from Eqs.~(\ref{conseqx2}) and (\ref{avB})
    \be\l{rhodisp4s}
    \Big\langle \left(\rho-n\right)^4 \Big\rangle \approx
    \Big\langle \left(\rho-n\right)^2 \Big\rangle^2\approx
    \frac{24 \langle \Delta^{(4)}_4\{\rho\}\rangle}{(\partial^4
      F/\partial \rho^4)^{}_{\rho=n}}~.
    \ee
    where $\Delta^{(4)}_4\{\rho\}$ is 
given by Eq.~(\ref{DF44s}),
    at the fourth order under the assumption of neglecting the
    second-order term. Notice that the angle brackets in Eq.~(\ref{rhodisp4s})
    have the same meaning as in Eqs.~(\ref{avsceq})-(\ref{correxp}).

    For the evaluation of average $\langle \Delta_4^{(4)}\rangle$
    in the last equation
    in (\ref{rhodisp4s}), with good accuracy within
    the mean field approximation, one can use the
    probability distribution $W_{\rm eq}^{(N)} \approx W^{(4)}_4$, 
      valid namely in the mean-field approximation,
    \be\l{W44s}
    W^{(4)}_4(\rho) =W^{(4),0}_4\exp\left[
      -\frac{F_4}{24T}(\rho-n)^4)\right]~,
    \ee
    where
    \be\l{norm44}
    W^{(4),0}_4=\left\{\int_0^\infty \mbox{d} \rho~\exp\left[
      -\frac{F_4}{24T}(\rho-n)^4)\right]\right\}^{-1};
    \ee
    see Eq.~(\ref{W4}) without the second-order term.
    Therefore, as
    in Appendix \ref{appD}, one has
    \bea\l{av44s}
    & \langle {\Delta}^{(4)}_4\{\rho\}\rangle\equiv \frac{F_4}{24}
    \langle(\rho-n)^4\rangle\nonumber\\
    &\approx \frac{F_4}{24} \int_0^\infty (\rho-n)^4 W^{(4)}_4\mbox{d}\rho~,
    \eea
    where $W^{(4)}_4$ is the normalized probability distribution 
    of the fourth order with zero second-order term, Eq.~(\ref{W44s})
    ($\int_0^\infty W^{(4)}_4\mbox{d}\rho=1$).
       Calculating now analytically integral in Eq.~(\ref{av44s}), 
      and comparing the result with  the
     expression on very right of Eq.~(\ref{rhodisp4s}), one obtains
    \be\l{difF44s}
   \langle\Delta^{(4)}_4\{\rho\}\rangle\approx \frac{T}{4}~. 
    \ee
    Differentiating over $\rho$ the relationship (\ref{p})
    between the
    pressure $P(\rho)$ and free energy $F(\rho)$ for a constant temperature $T$,
     similarly as for the second order case, one can express
the fourth derivative
of $F(\rho)$ over $\rho$ at $\rho=n$ in terms of the
second derivative of the incompressibility $\mathcal{K}$,
\bea\l{dFK44s}
&\left(\frac{\partial^4 F(\rho)}{\partial \rho^4}\right)_{\rho=n}=
\frac{\langle N \rangle\mathcal{K}^{\prime\prime}(n)}{n^2},~~~ \mbox{with}
\nonumber\\
&\mathcal{K}^{\prime\prime}(n)=
\left(\frac{\partial^3 P(\rho)}{\partial \rho^3}\right)_{\rho=n}~,
\eea
where $P$ is the pressure, $P(T,\rho)$, Eq.~(\ref{p}),
and $P=P(T,n)$ is the equation of state in canonical variables.
Using Eqs.~(\ref{difF44s}) and (\ref{dFK44s}),
from the particle number density dispersion $\mathcal{D}_\rho$,
normalized by $n^2$, Eq.~(\ref{FL-MM}), at the fourth-order expansion
of the free energy (taking again zero for the second-order term),
$\mathcal{D}^{(4)}_4$, 
with the probability distribution $W^{(4)}_4$,
Eq.~(\ref{W44s}), i.e.,
in the mean field (zero-order correlations)
approximation, 
one naturally obtains
the same limit as given in Eq.~(\ref{D4limCP}).
Employing finally the same normalization of the dispersion $D_N$
by $\langle N\rangle$, in order to compare with Eq.~(\ref{FL-press}),
we arrive at the expression (\ref{lim4}), derived early in
Ref.~\cite{RJ58}.

\renewcommand{\theequation}{F.\arabic{equation}}
\renewcommand{\thesubsection}{F\arabic{subsection}}
  \setcounter{equation}{0}

  \section{Other improved approach to the particle number fluctuations}
\l{appF}

Following Ref.~\cite{TK66} we assume in fact the mean field approximation
  $W_4(\rho)$, Eq.~(\ref{W4}), for the Gibbs distribution averaged in the
  phase space and particle numbers
from the very
beginning, everywhere in the calculations of particle number fluctuations.
Finally, for
calculations of the dispersion (variance) 
$D_{\rho}=\langle(\rho-n)^2\rangle$, one obtains
\be\l{Drhom}
D_\rho=\langle (\rho-n)^2\rangle=\mathcal{M}_2/\mathcal{M}_0~,
\ee
where
\bea\l{m}
&\mathcal{M}_m(c^{}_2,c^{}_4)=
\int_0^\infty  \mbox{d}\rho~(\rho-n)^m\nonumber\\
&\times\exp\left[
  -c^{}_2 \left(\rho-n\right)^2-c^{}_4\left(\rho-n\right)^4\right]\nonumber\\
&\approx
2\int_0^\infty \mbox{d}x~ x^{m} \nonumber\\
&\times\exp\left(-c^{}_2x^2-c^{}_4x^4\right)~,
\quad m=0,~2~,
\eea
\be\l{cm}
c^{}_2=\frac{F_2}{2T}~,\quad  c^{}_4=\frac{F_4}{24T}~;
\ee
see Eq.~(\ref{Fm}) for the derivatives $F_m$ of the free energy $F$. From
Eq.~(\ref{m}) one obtains the explicit expressions for the moments
of the distribution function, $\mathcal{M}_m$, in terms of the MacDonald
Bessel functions $K_\nu(\alpha/8)$,
\bea\l{MK}
& \mathcal{M}_2=\frac{1}{8}\left(c^{}_2/c^{}_4\right)^{3/2}
\exp\left(-\alpha/8\right)\nonumber\\
&\times\left[K_{3/4}(\alpha/8)-K_{1/4}(\alpha/8)\right]~,\nonumber\\
& \mathcal{M}_0=\frac{1}{2}\left(c^{}_2/c^{}_4\right)^{1/2}
\exp\left(\alpha/8\right)
K_{1/4}(\alpha/8)~,
\eea
where 
\be\l{alpha}
\alpha=c_2^2/c^{}_4
\ee
[see Eqs.~(\ref{alp}) and (\ref{cm})].

\subsection{The limit case $c^{}_2=0$}

Taking the limit $c^{}_2 \rightarrow 0$ to the critical point,
from Eqs.~(\ref{Drhom}) with
Eq.~(\ref{MK}) for the moments $\mathcal{M}_m$, one obtains the dispersion:
\bea\l{D0c2}
&\mathcal{D}_\rho=\frac{\Gamma(3/4)}{\Gamma(1/4)}~\frac{1}{\sqrt{c^{}_4}}=
0.338 \sqrt{\frac{24T}{F_4}}\nonumber\\
&=
0.676\sqrt{\frac{6Tn^2}{\langle N\rangle \mathcal{K}^{\prime\prime}}}~.
\eea
For the normalized dispersion, $D_\rho/n^2$, one finally finds
\be\l{D0c2n2}
\frac{\mathcal{D}_\rho}{n^2}=
0.676\sqrt{\frac{6T}{\langle N\rangle n^2 \mathcal{K}^{\prime\prime}}}~.
\ee
The constant in front of the square root is smaller than that in
Eq.~(\ref{D4limCP}).
For the particle number fluctuation $\omega$, Eq.~(\ref{omgenN}),
at the critical point, from Eq.~(\ref{D0c2}) one
approximately finds
\be\l{om0c2}
\omega=1.66\sqrt{\frac{T\langle N \rangle}{n^2 \mathcal{K}^{\prime\prime}}}~.
\ee

\subsection{The limit case $c^{}_4=0$}

Taking the limit $c^{}_4 \rightarrow 0$,
from Eqs.~(\ref{Drhom}) with
Eq.~(\ref{MK}) for the moments $\mathcal{M}_m$, one obtains
\be\l{D0c4}
D_\rho=\frac{1}{2 c^{}_2}~\frac{T}{F_2}=
\frac{Tn^2}{\langle N \rangle \mathcal{K}}~.
\ee
Similarly as in the previous subsection of this appendix,
for the particle number fluctuation $\omega$, Eq.~(\ref{omgenN}),
 from Eq.~(\ref{D0c4}) one
approximately finds
\be\l{om0c4}
\omega=T/\mathcal{K}~.
\ee
%



\references

  \bibitem{bethe} H.A.~Bethe, Rev. Mod. Phys. {\bf 9}, 69 (1937); {\it Theory of Nuclear Matter},
    Annu. Rev. Nucl. Sci. {\bf 21}, 93 (1971).

\bibitem{migdal} A.B.~Migdal, {\it The Finite Fermi-System Theory and
Properties of Atomic Nuclei} (Interscience, New York,
1967; Nauka, Moscow, 1983).

\bibitem{MS69} W.D.~Myers and W.J. Swiatecki, Ann. Phys. (N.Y.)
  {\bf 55}, 395 (1969); {\bf 84}, 186 (1974).

\bibitem{BD72} M.~Brack, J.~Damgard, A.S.~Jensen et al., Rev. Mod.
  Phys. {\bf 44}, 320 (1972).

 \bibitem{RS80} P.~Ring and P.~Schuck, {\it
  The Nuclear Many-Body Problem} (Springer-Verlag, New York,
   Heidelberg, Berlin, 1980).

\bibitem{BG85} M.~Brack, C.~Guet, and H-B.~H{\aa}kansson, Phys. Rep. {\bf 123}, 275 (1985).  

\bibitem{BHR03} M.\ Bender, P.H.\ Heenen, and P.G.\ Reinhard, Rev. Mod. Phys. {\bf 75}, 121 (2003).

 \bibitem{KS20} V.M.\ Kolomietz and S.Shlomo, {\it
   Mean Field Theory} (World Scientific, 2020).

\bibitem{LLv5} L.D.\  Landau and E.M.\ Lifshitz,  {\it Statistical Physics,
 Course of
Theoretical Physics} (Pergamon, Oxford, UK, 1975), Vol.~5.     

\bibitem{huang} K.~Huang, {\it Statistical Mechanics} (Wiley \& Sons,
  New York,
    1963, 1st edition; 1987, 2nd edition).  

  \bibitem{AB00} A.~Bonasera, M.~Bruno, C.O.~Dorso, and P.F.~Mastinu,  Riv. Nuovo Cimento, {\bf 23}, 1 (2000).

  \bibitem{AB08} A.~Bonasera, Z.~Chen, R.~Wada, K~Hagel, J.~Natowitz, P.~Sahu, L.~Qin, S.~Kowalski,
    Th.~Keutgen, T.~Materna, and T.~Nakagawa, Phys. Rev. Lett, {\bf 101}, 122702 (2008).

    \bibitem{AB10} M.~Huang, A.~Bonasera, Z.~Chen, R.~Wada, K.~Hagel, J.B.~Natowitz, P.K.~Sahu, L.~Qin, T.~Keutgen,
    S.~Kowalski, T.~Materna, J.~Wang, M.~Barbui, C.~Bottosso, and M.R.D.~Rodrigues, Phys. Rev. C {\bf 81}, 044618 (2010).
  
    \bibitem{ex-1} J.E.~Finn, S.~Agarwal, A.~Bujak, J.~Chuang, L.J.~Gutay,
      A.S.~Hirsch, R.W.~Minich, N.T.~Porile, R.P.~Scharenberg,
      B.C.~Stringfellow, and F.~Turkot, Phys. Rev. Lett. {\bf 49}, 1321 (1982).

\bibitem{ex-2} R.W.~Minich et al., Phys. Lett. B {\bf 118}, 458 (1982).

\bibitem{ex-3}  A.S.~Hirsch, A.~Bujak, J.E.~Finn, L.J.~Gutay, R.W.~Minich,
  N.T.~Porile, R.P.~Scharenberg, B.C.~Stringfellow, and F.~Turkot,
  Phys. Rev. C {\bf 29}, 508 (1984).

\bibitem{ex-4} J.~Pochodzalla, T.~Mohlenkamp, T.~Rubehn, A.~Schuttauf,
  A.~Worner, E.~Zude et al., Phys. Rev. Lett. {\bf 75}, 1040 (1995).

\bibitem{ex-5} J.B.~Natowitz, K.~Hagel, Y.~Ma, M.~Murray, L.~Qin, R.~Wada, and
  J.~Wang, Phys. Rev. Lett. {\bf 89}, 212701 (2002).

\bibitem{ex-5a} J.B.~Natowitz, R.~Wada, K.~Hagel, T.~Keutgen, M.~Murray,
  A.~Makeev, L.~Qin,
  P.~Smith, and C.~Hamilton, Phys. Rev. C {\bf 65}, 034618 (2002).

\bibitem{ex-6} V.A.\ Karnaukhov, H.\ Oeschler, S.P.\ Avdeyev, E.V.\ Duginova,
  V.K.\ Rodionov, A.\ Budzanowski, W.\ Karcz, O.V.\ Bochkarev, E.A.\ Kuzmin,
  L.V.\ Chulkov, E.\ Norbeck, and A.S.\ Botvina, Phys. Rev. C {\bf 67},
  011601(R) (2003).

\bibitem{vova}
  V.\ Vovchenko,  A.\ Motornenko, P.\ Alba, M.I.\ Gorenstein, L.M.\ Satarov,
  and H.\ Stoecker,
  Phys. Rev. C {\bf 96}, 045202 (2017).

  \bibitem{satarov} L.M.\ Satarov, I.N.\ Mishustin, A.\ Motornenko, V.\ Vovchenko,
    M.I.\ Gorenstein, and H.\ Stoecker, Phys. Rev. C {\bf 99},
  024909 (2019).

\bibitem{roma1} R.V.\ Poberezhnyuk, V.\ Vovchenko, M.I.\ Gorenstein,
  and H.\ Stoecker,
  Phys. Rev. C {\bf 99}, 024907 (2019).

  \bibitem{marik} V.\ Vovchenko, D.V.\ Anchishkin, and M.I.\ Gorenstein, Phys. Rev. C
  {\bf 91}, 0.64314 (2015).

\bibitem{roma} R.V.\ Poberezhnyuk, V.\ Vovchenko, D.V.\ Anchishkin, and
  M.I.\ Gorenstein,
    J. Mod. Phys. E {\bf 26}, 1750061 (2017).

 \bibitem{satarov1} L.M.\ Satarov, M.I.\ Gorenstein, I.N.\ Mishustin, and
   H.\ Stoecker, Phys. Rev. C {\bf 101}, 024913 (2020).

 \bibitem{roma2} R.V.\ Poberezhnyuk, O.\ Savchuk, M.I.\ Gorenstein,
   V.\ Vovchenko,
   K.\ Taradiy, V.V.\ Begun, L.\ Satarov, J.\ Steinheimer, and H.\ Stoecker,
   Phys. Rev. C {\bf 102},
   024908 (2020).
   
\bibitem{oleh20} O.\ Savchuk, Y.\ Bondar, O.\ Stashko, R.V.\ Poberezhnyuk,
  V.\ Vovchenko, M.I.\ Gorenstein, and H.\ Stoecker, Phys. Rev. C {\bf 102},
  035202 (2020).

\bibitem{St21-1} O.S.\ Stashko, D.V.\ Anchishkin, O.V.\ Savchuk, and
  M.I.\ Gorenstein, J. Phys. G {\bf 48}, 055106 (2021). 

\bibitem{St21-2}  O.S.\ Stashko, O.V.\ Savchuk,
  R.V.\ Poberezhnyuk, V.\ Vovchenko, and M.I.\ Gorenstein,
  Phys. Rev. C {\bf 103}, 065201 (2021).

     \bibitem{Kuzn21}
     V.A.\ Kuznietsov, O.S.\ Stashko, O.V.\ Savchuk,
     and M.I.\ Gorenstein, Phys. Rev. C {\bf 104}, 055202 (2021).     

\bibitem{AC90} M.\ Anisimov and V.\ Sychev,
  {\it Thermodynamics of critical state  for individual sustances},
  (Energoatomizdat, Moscow, 1990)(in Russian).

\bibitem{KG67} L.P.\ Kadanoff, W.\ G\"otze, D.\ Hamblen, R.\ Hecht, E.A.S.\ Levis, V.V.\ Palciaukas,
  M.\ Rayl, and J.\ Swift, Rev. Mod. Phys. {\bf 39}, 395 (1967).

\bibitem{An87} M.A.\ Anisimov, {\it Modern Physics Problems,
  Critical phenomena in liquids and liquid cristals},
  {Fiz.-mat.-lit., ``Nauka'', Moscow, 1987).

\bibitem{Sa99} M.V.\ Sadovsky, {\it Lectures on the statistical physics}, Jekaterinburg,
  Institute of Electrophysics UrO Russian Academy of Science, Russia, (1999). 

\bibitem{ZR88} L.A.\ Zubkov and V.P.\ Romanov, Usp. Phys. Sci., {\bf 14} (no 4), 615 (1988).  

 \bibitem{BR75} R.\ Balescu, {\it Equilibrium and nonequilibrium statistical
   mechanics} (Wiley, New York, 1975), Vol.~1.

  \bibitem{TR38} R.C.\ Tolman,  {\it The principles of statistical mechanics}
  (Oxford at the Clarendon Press, Oxford, 1938).

\bibitem{RJ58} J.S.\ Rowlinson,{\it The properties of real gases},
  Encyclopedia of Physics,
  (Springer-Verlag, Academic Edition, Berlin, 1958), Vol. 3/12,
  ISBN : 978-3-642-45894-1.

\bibitem{TK66} K.B.~Tolpygo, {\it Thermodynamics
and Statistical Physics} (Kiev University, Kiev, 1966) (in Russian).

\bibitem{IA71} A.~Ishihara, {\it Statistical Physics}
  (Academic Press, New York, 1971).
  
\bibitem{ZM02} D.\ Zubarev, V.\ Morozov, and G.\ R\'opke, 
  {\it Statistical Mechanics of Nonequilibrium Processes},
  (Moscow, Fizmatlit, 2002)(in Russian), Vol.~1.

\bibitem{KW04} A.\ Kostrowicka Wyczalkowska, J.V.\ Senders, and M.A.\ Anisimov,
  Physica A: Statistical Mechanics and its Applications, {\bf 334}, 482 (2004).

\bibitem{FMG19} S.N.\ Fedotkin, A.G.\ Magner, and M.I.\ Gorenstein,
  Phys. Rev. C {\bf 100}, 054334 (2019).  

   \bibitem{FMG22}  S.N.\ Fedotkin, A.G.\ Magner, and U.V.\ Grygoriev,
     Phys. Rev. C {\bf 105}, 024621 (2022).

 \bibitem{AV-15} D.~Anchishkin and V.~Vovchenko,
   J. Phys. G {\bf 42}, 105102 (2015).

 \bibitem{satarov0} L.M.\ Satarov, M.I.\ Gorenstein, A.\ Motornenko,
   V.\ Vovchenko,
    I.N.\ Mishustin, and H.\ Stoecker, J. Phys. G {\bf 44}, 125102 (2017). 

  \bibitem{La81} J.M.\ Lattimer, Ann. Rev. Nucl. Phys. Part. Sci.
    {\bf 31}, 337 (1981); ibid {\bf 62}, 425 (2012).

     \bibitem{VGS-17}
  V.\ Vovchenko, M.I.\ Gorenstein, and H.\ Stoecker,
  Phys. Rev. Lett. {\bf 118}, 182301 (2017).
  
 \bibitem{VJGS-18}   V.\ Vovchenko, L.\ Jiang, M. I.\ Gorenstein, and
   H.\ Stoecker, Phys. Rev. C {\bf 98}, 024910 (2018).

   \bibitem{FMG20}  S.N.\ Fedotkin, A.G.\ Magner, and U.V.\ Grygoriev,
    arXiv:2012.09695 v2 [nucl-th], 2020.  

     \bibitem{Sm08} M.v.~Smoluchowski, Ann. d. Phys. {\bf 330}, 205 (1908).

     \bibitem{Ei10} A.~Einstein, Annalen der Physik, {\bf 338}, 1275 (1910).   

  \bibitem{MS21npa}  A.G.\ Magner, A.I.\ Sanzhur, S.N.\ Fedotkin,
   A.I.\ Levon, and S.\ Shlomo,
       Nucl. Phys. A {\bf 1021}, 122423 (2022).

\bibitem{MS21prc}  A.G.\ Magner, A.I.\ Sanzhur, S.N.\ Fedotkin, A.I.\ Levon, and S.\ Shlomo,
  Phys. Rev. C {\bf 104}, 044319 (2021).

\bibitem{MS21ijmpe}  A.G.\ Magner, A.I.\ Sanzhur, S.N.\ Fedotkin,
  A.I.\ Levon, and S.\ Shlomo,
   Int. J. Mod. Phys. E {\bf 30}, 2150092 (2021).

\bibitem{MS22FLT}  A.G.\ Magner, A.I.\ Sanzhur, S.N.\ Fedotkin,
  A.I.\ Levon, U.V.\ Grygoriev, and S.\ Shlomo,
  Low Temperature Physics, {\bf 48}, 920 (2022).

  \bibitem{PercolinMod} D.~Stauffer and A.~Aharony, Introduction to
  Percolation Theory (Taylor and Francis,
London) 1992.  

\bibitem{mafm}
 A.G.\ Magner, K.\ Arita, S.N.\ Fedotkin, and K.\ Matsuyanagi,
 Prog. Theor. Phys. {\bf 108}, 853 (2002).

 \bibitem{MY11} 
A.G.\ Magner, Y.S.\ Yatsyshyn, K.\ Arita, and M.\ Brack,  Phys. At. Nucl.
{\bf 74}, 1445 (2011).

\bibitem{maf}  A.G.\ Magner, K.\ Arita, and S.N.\ Fedotkin,
  Progr. Theor. Phys. {\bf 115}, 523 (2006).

\bibitem{Fe62}
 M.V.\ Fedoriuk,
 Sov. J. of Comput. Math. Math. Phys. {\bf 2}, 145 (1962); ibid {\bf 4}, 671
 (1964).

\bibitem{Fe77} M.V.\ Fedoriuk,  {\it The method of steepest descents}
  (Nauka, Moscow, 1977) (in Russian). 

  \bibitem{MK16} A.G.\ Magner, M.V.\ Koliesnik, and K.\ Arita, Phys. At. Nucl.
  {\bf 79}, 1067 (2016).

\bibitem{MA17} A.G.\ Magner and K.\ Arita, Phys. Rev. E {\bf 96}, 042206 (2017).
\end{document}